\documentclass[iop]{emulateapj}
\usepackage[english]{babel}
\usepackage{multirow}
\usepackage{gensymb}
\usepackage{amsmath, amssymb}
\usepackage{booktabs}
\newcommand{\hb}{H$\beta$}
\newcommand{\ha}{H$\alpha$}
\newcommand{\lya}{Ly$\alpha$}
\newcommand{\oiii}{[\ion{O}{3}]}

\newcommand{\esc}{erg cm$^{-2}$ s$^{-1}$}
\newcommand{\es}{erg s$^{-1}$}

\newcommand{\kms}{km s$^{-1}$}
\newcommand{\myr}{M$_{\odot}$ yr$^{-1}$}
\newcommand{\msun}{M$_{\odot}$}

\begin{document}
\title{A giant L\lowercase{y}$\alpha$ nebula in the core of an X-ray
  cluster at $\lowercase{z}=1.99$:\\ implications for early energy injection}
\author{\sc Francesco Valentino \altaffilmark{1,2}}
\altaffiltext{1}{Laboratoire AIM-Paris-Saclay, CEA/DSM-CNRS-Universit\'{e} Paris Diderot, Irfu/Service d'Astrophysique, CEA Saclay, Orme des Merisiers, F-91191 Gif sur Yvette, France}
\altaffiltext{2}{francesco.valentino@cea.fr}
\author{\sc Emanuele Daddi\altaffilmark{1}}
\author{\sc Alexis Finoguenov\altaffilmark{3,4}}
\altaffiltext{3}{Department of Physics, University of Helsinki, Gustaf
Ha\"{a}llstr\"{o}min katu 2a, 00014 Helsinki, Finland}
\altaffiltext{4}{Center for Space Sciences and Technology, University
  of Maryland, Baltimore County, 1000 Hilltop cir., Baltimore, MD 21250, USA}
\author{\sc Veronica Strazzullo\altaffilmark{1,5}}
\altaffiltext{5}{Department of Physics, Ludwig-Maximilians-Universit\"{a}t, Scheinerstr. 1, 81679 M\"{u}nchen, Germany}
\author{\sc Amandine Le Brun\altaffilmark{1}}
\author{\sc Cristian Vignali\altaffilmark{6,7}}
\altaffiltext{6}{Dipartimento di Fisica e Astronomia, Universit\`a degli Studi di
Bologna, Viale Berti Pichat 6/2, 40127 Bologna, Italy}
\altaffiltext{7}{INAF – Osservatorio Astronomico di Bologna, Via Ranzani 1,
40127 Bologna, Italy}
\author{\sc Fr\'ed\'eric Bournaud\altaffilmark{1}}
\author{\sc Mark Dickinson\altaffilmark{8}}
\altaffiltext{8}{National Optical Astronomy Observatories, 950 N Cherry Avenue, Tucson, AZ 85719, USA}
\author{\sc Alvio Renzini\altaffilmark{9}}
\altaffiltext{9}{INAF-Osservatorio Astronomico di Padova Vicolo dell'Osservatorio 5, I-35122 Padova, Italy}
\author{\sc Matthieu B\'{e}thermin\altaffilmark{10}}
\altaffiltext{10}{European Southern Observatory, Karl-Schwarzschild-Str. 2, 85748 Garching, Germany}
\author{\sc Anita Zanella\altaffilmark{1}}
\author{\sc Rapha\"{e}l Gobat\altaffilmark{1,11}}
\altaffiltext{11}{School of Physics, Korea Institute for Advanced Study, Hoegiro 85, Dongdaemun-gu, Seoul 02455, Republic of Korea}
\author{\sc Andrea Cimatti\altaffilmark{6,7}}
\author{\sc David Elbaz\altaffilmark{1}}
\author{\sc Masato Onodera\altaffilmark{12,13}}
\altaffiltext{12}{Institute for Astronomy, ETH Z\"{u}rich Wolfgang-Pauli-strasse 27, 8093 Z\"{u}rich, Switzerland}
\altaffiltext{13}{Subaru Telescope, National Astronomical Observatory of Japan 650 North A'ohoku Place, Hilo, HI 96720, USA}
\author{\sc Maurilio Pannella\altaffilmark{1,5}}
\author{\sc Mark Sargent\altaffilmark{14}}
\altaffiltext{14}{Astronomy Centre, Department of Physics and Astronomy,
University of Sussex, Brighton, BN1 9QH, UK}
\author{\sc Nobuo Arimoto\altaffilmark{13,15}}
\altaffiltext{15}{Graduate University for Advanced Studies, 2-21-1 Osawa, Mitaka, Tokyo, Japan}
\author{\sc Marcella Carollo\altaffilmark{12}}
\author{\sc Jean-Luc Starck\altaffilmark{1}}

\begin{abstract}
We present the discovery of a giant $\gtrsim$100~kpc Ly$\alpha$ nebula
detected in the core of the X-ray emitting cluster CL~J1449+0856 at
$z=1.99$ through Keck/LRIS narrow-band imaging. This detection extends the known relation between \lya\
nebulae and overdense regions of the Universe to the dense core of a
$5-7\times10^{13}$ \msun\ cluster. The most plausible candidates to
power the nebula are two Chandra-detected AGN host
cluster members, while cooling from the X-ray phase and cosmological cold flows are 
disfavored primarily because of the high
\lya\ to X-ray luminosity ratio
($L_{\mathrm{Ly\alpha}}/L_{\mathrm{X}} \approx0.3$,
$\gtrsim10-1000\times$ higher than in local cool-core clusters) and by
current modeling.
Given the physical conditions of the \lya-emitting gas and the possible interplay with the X-ray phase, we
argue that the \lya\ nebula would be short-lived ($\lesssim10$ Myr)
if not continuously replenished with cold gas at a rate of
$\gtrsim1000$ \myr. We investigate the possibility that cluster galaxies supply
  the required gas through outflows and we show that 
  their total mass outflow rate matches the replenishment necessary to sustain the nebula.
This scenario directly implies the extraction of energy
from galaxies and its deposition in the surrounding intracluster
medium, as required to explain the thermodynamic properties of local
clusters. We estimate an energy injection of the order of
$\thickapprox2$ keV per particle in the intracluster medium over a
$2$ Gyr interval. In our baseline calculation AGN provide up to $85$\% of the
injected energy and 2/3 of the
mass, while the rest is supplied by supernovae-driven winds. 
\end{abstract}	


\section{Introduction}
\label{sec:introduction}
Since their first
discovery in the late 1990s \citep{francis_1996, steidel_2000},
high-redshift, extended ($\gtrsim100$ kpc), and luminous (few
$10^{43}-10^{44}$ \es) gas reservoirs shining by the emission of \lya\
photons have progressively become a matter of debate. Despite two
decades of investigation, several
aspects of these ``\lya\ nebulae'' remain puzzling, including the
origin of the \lya-emitting gas, its powering mechanism,
the possible effects on the evolution of the embedded galaxies,
and, ultimately, its fate \citep[i.e.,][]{matsuda_2004, dey_2005,
  geach_2009, prescott_2009, cantalupo_2014}.
Understanding where \lya\ nebulae fit in the current theoretical framework of structure
formation has sparked particular interest, since they call into
question a cornerstone of modern astrophysics: the complex interplay of supply,
consumption, and expulsion of gas that shapes high-redshift systems.
In this work we focus on a specific feature of \lya\ nebulae: the
connection with their surrounding environment. This perspective
complements the approaches already presented in the literature and
allows us to shed light on several of the problematics listed above. 
First, there are observational hints that \lya\ nebulae preferentially reside in
overdense regions of the Universe or sparse protoclusters
\citep{steidel_2000, matsuda_2004, venemans_2007}. This suggests a
possible connection with the formation of massive structures, even if 
it is not clear in which density regimes this correlation holds. 
Interestingly, in the local Universe the presence of kpc-size, filamentary reservoirs of
ionized gas in the center of ``cool-core'' X-ray emitting
clusters (CCs) has been known for decades
\citep{fabian_1984, heckman_1989, hatch_2007, mcdonald_2010,
  tremblay_2015}. From this angle, it is tempting to view the
high-redshift \lya\ nebulae as the counterparts of
local filaments, with sizes and luminosities reflecting the extreme
conditions of the primordial Universe \citep{mcdonald_2010, arrigonibattaia_2015}. However,
the detailed physics of the nebular emission is still debated even for
local clusters, despite the quality of the available data. A mix of different
heating mechanisms is probably at the origin of the emission by the ionized filaments,
with a possible important role played by young stars formed
\textit{in-situ} (\citealt{tremblay_2015} and references therein). The origin of
the cold gas has not been fully clarified either: even if
modern models of auto-regulated cooling from the X-ray emitting
intracluster medium (ICM) successfully
reproduce several properties of the nebulae in CCs
\citep[i.e.,][]{gaspari_2012, voit_2015, tremblay_2015}, the cold gas
might also originate from a starburst event or the
active galactic nuclei (AGN) in the central brightest cluster galaxy \citep[BCG,][]{hatch_2007},
or be uplifted by propagating radio-jets and buoyant X-ray bubbles
\citep{churazov_2001, fabian_2003}, or stripped in a
recent merger \citep{bayer-kim_2002, wilman_2006}. 
Therefore physical insights might not be straightforwardly gained from
the simple observation of local filaments.\\
An attempt at assessing the validity of this suggestion can be done
through the observation of giant \lya\ nebulae in the core of high-redshift
galaxy clusters. To date, we have lacked strong observational evidence
primarily because of the scarcity of 
\textit{bona fide} X-ray emitting structures discovered
at $z\geq1.5$ \citep[i.e.,][]{andreon_2009, papovich_2010,
  santos_2011, stanford_2012,
  gobat_2011, gobat_2013, brodwin_2015}. Here we study in detail the
case of the most distant among these X-ray detected structures,
CL~J1449+0856 at $z=1.99$ \citep[G11, G13 and S13
hereafter]{gobat_2011, gobat_2013, strazzullo_2013}. Its 
extended emission from hot plasma and the dominant population of red,
massive, and passive galaxies in the compact core (G11, G13, S13) place it in a more
advanced evolutionary stage than protoclusters at
similar redshift and make it a suitable candidate to start the search
for nebulae in far away clusters. In this work we present the results
of a recent narrow-band imaging campaign we conducted with Keck/LRIS,
with which we identified a $\sim100$~kpc \lya-emitting nebula in the
cluster core. However, the detailed analysis of the conditions of the
nebula and its environment shows some tensions with the current
picture of filaments in local clusters. Even if cooling from the X-ray
emitting plasma may partially contribute to the \lya\ luminosity, the
nebula is plausibly powered by AGN in the
cluster core.\\
Motivated by this discovery, we further investigate the relationship
between the \lya\ nebula, galaxy activity in the form of star
formation and black hole growth, and the total energy content of the
ICM at this early stage of the cluster evolution. The latter is a
controversial issue in modern astrophysics. In fact, it has been known for more than
two decades that the observed X-ray properties of the ICM in nearby
clusters are inconsistent with the predictions from pure gravitational settling
and an extra energy contribution is missing \citep{kaiser_1991,
ponman_1999, tozzi_2001}. In cosmological simulations
this energy is provided by star-forming galaxies (SFGs) and AGN
through outflows, and their efficiencies can be
calibrated to reproduce the properties of the local Universe
\citep[e.g.,][]{lebrun_2014, pike_2014}.
Although the most successful models are those in which
heating of the ICM happens early, such as (cosmo-)OWLS
\citep{schaye_2010, mccarthy_2011}, this process is
still poorly constrained observationally: the timing and duration of
this phenomenon, its main energy source (galactic winds from
either supernovae or AGN), and the energy
transfer mechanism are subject to debate (i.e., \citealt{mccarthy_2011,
dave_2008, mcnamara_2007, fabian_2012}). Here we argue that 
the presence of the \lya\ nebula is interlaced with the observed vigorous activity of
galaxies in the cluster core and that it may signpost a significant energy
injection into the ICM. Eventually, we estimate the amount of this injection due to strong galaxy feedback during a phase that, if
prevalent in high-redshift 
structures, would be crucial to set the final energy budget and metal
content of present-day clusters.\\ 
This paper is organized as follows: in Section \ref{sec:data} we present
the narrow- and broad-band imaging observations that led to the
discovery of the \lya\ nebula, along with the results of a recent
Chandra follow-up of CL~J1449+0856; in Section \ref{sec:physics} we
discuss the physical properties of the \lya\ nebula, the possible
powering mechanisms, and the timescales regulating its evolution,
concluding that a substantial gas replenishment is necessary to feed the system. In
Section \ref{sec:discussion} we focus on galaxy outflows as a
plausible source of gas replenishment and we study the corresponding
injection of energy into the ICM. Concluding
remarks are presented in Section \ref{sec:conclusions}. Unless
stated otherwise, we assume a $\Lambda$CDM cosmology with
$\Omega_{\rm m} = 0.3$, $\Omega_{\rm \Lambda} = 0.7$, and $H_0 = 70$
km s$^{-1}$ Mpc$^{-1}$ and a Salpeter initial mass function (Salpeter
1955). All magnitudes are expressed in the AB system.  
\begin{figure}
  \centering
  \includegraphics[width=\columnwidth]{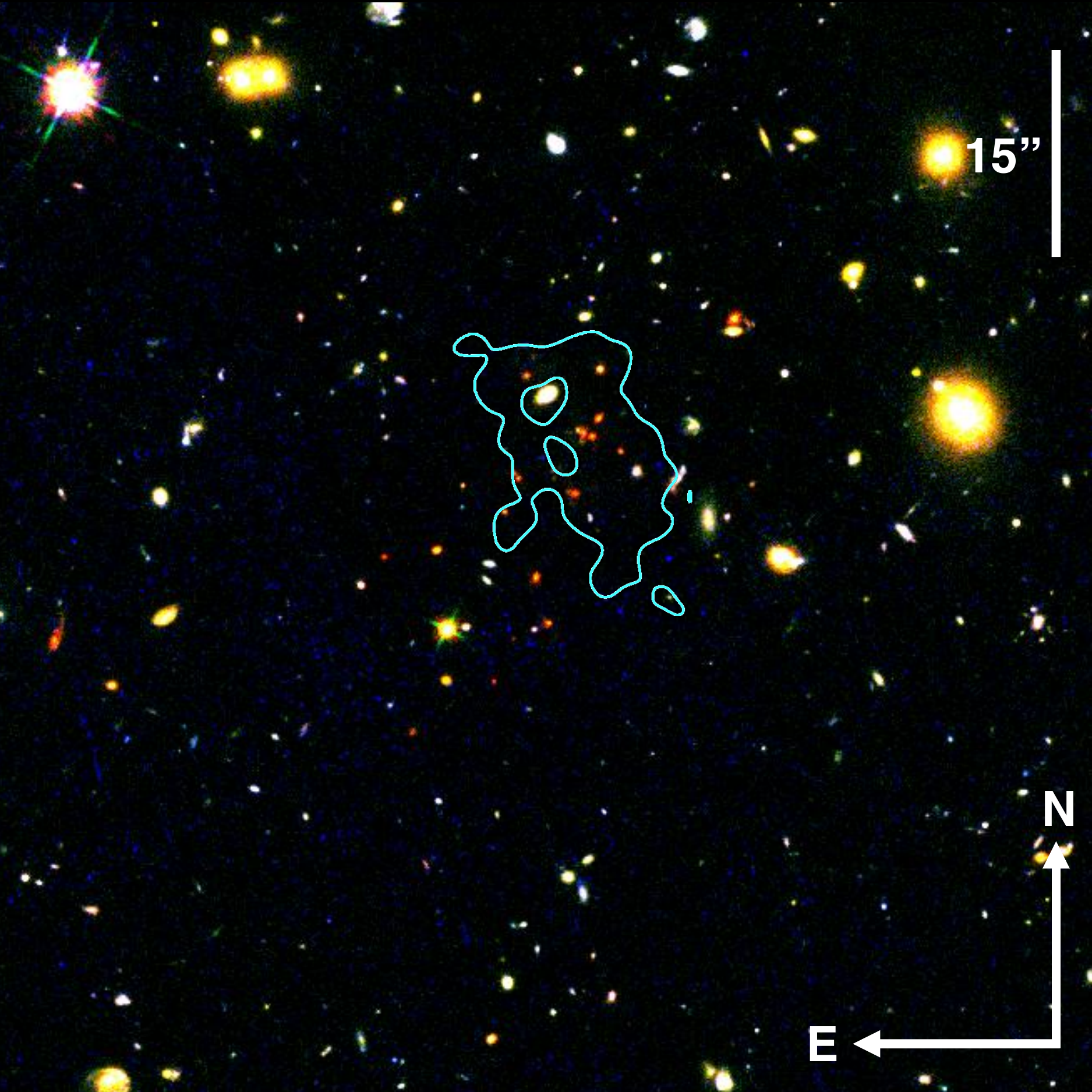}
  \caption{\textbf{Cluster of galaxies CL~J1449+0856 at $z=1.99$.}
    HST/F140W (red), F105W (green), and F606W (blue) RGB-composite
    image of CL~J1449+0856. The central concentration of red galaxies
    represents the core of the cluster. The cyan line marks the
    1$\sigma$ contour of the \lya\ nebula from the wavelet reconstruction.} 
  \label{fig:rgb}
\end{figure}

\section{Observations and data analysis}
\label{sec:data}
In this section we present the Keck/LRIS narrow-band imaging of
CL~1449+0856. We also describe recent Chandra observations which we use to
update the cluster X-ray properties previously constrained by
XMM-Newton follow-up only. Specifically, we revise the total extended X-ray
luminosity, gas temperature, and halo mass, presenting a new estimate
from the velocity dispersion.

\subsection{The \lya\ nebula detection: narrow-band imaging}
\label{sec:nb}
\begin{figure*}
  \centering
  \includegraphics[width=0.75\textwidth]{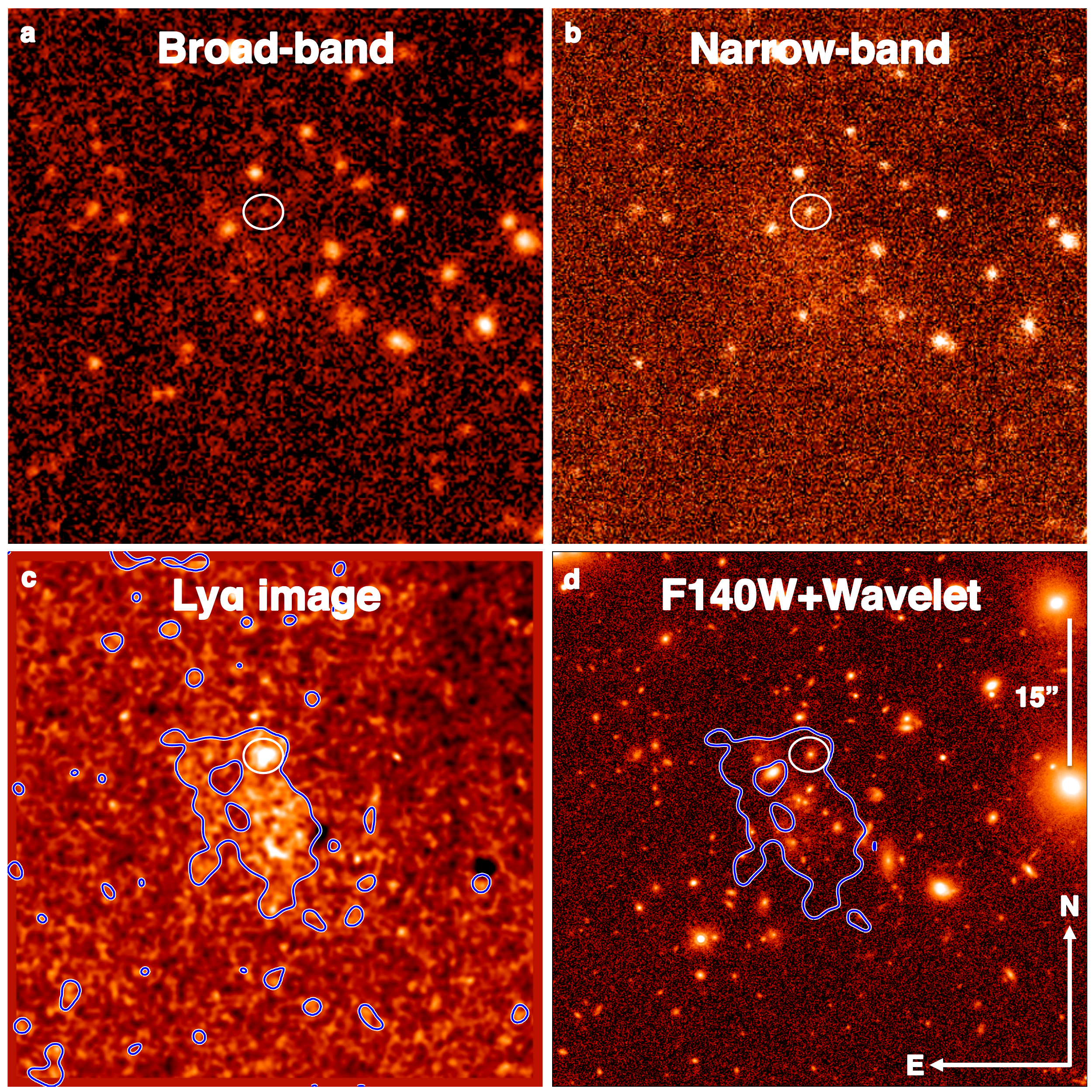}
  \caption{\textbf{100-kpc extended \lya\ nebula at $z=1.99$.}
    Images of CL~J1449+0856 in the broad \textit{U} band (panel
    \textbf{a}) and NB3640 narrow-band (panel \textbf{b}), and a
    continuum-subtracted \lya\ emission line map
    smoothed on scales of 1" (panel \textbf{c}). The white
    circle indicates the heavily obscured
    AGN \#661 (G13). The extended emission southward
    is the \lya\ nebula. Panel \textbf{d} shows the HST/WFC3 F140W
    image. In panels \textbf{c} and \textbf{d} the blue line marks
    the 1$\sigma$ contour of the large scale \lya\ emission from the
    wavelet reconstruction after the subtraction of point-like
    sources. For reference, 15'' correspond to $\sim125$~kpc at $z=1.99$.}
  \label{fig:images}
\end{figure*}

\begin{figure}
  \centering
  \includegraphics[width=\columnwidth]{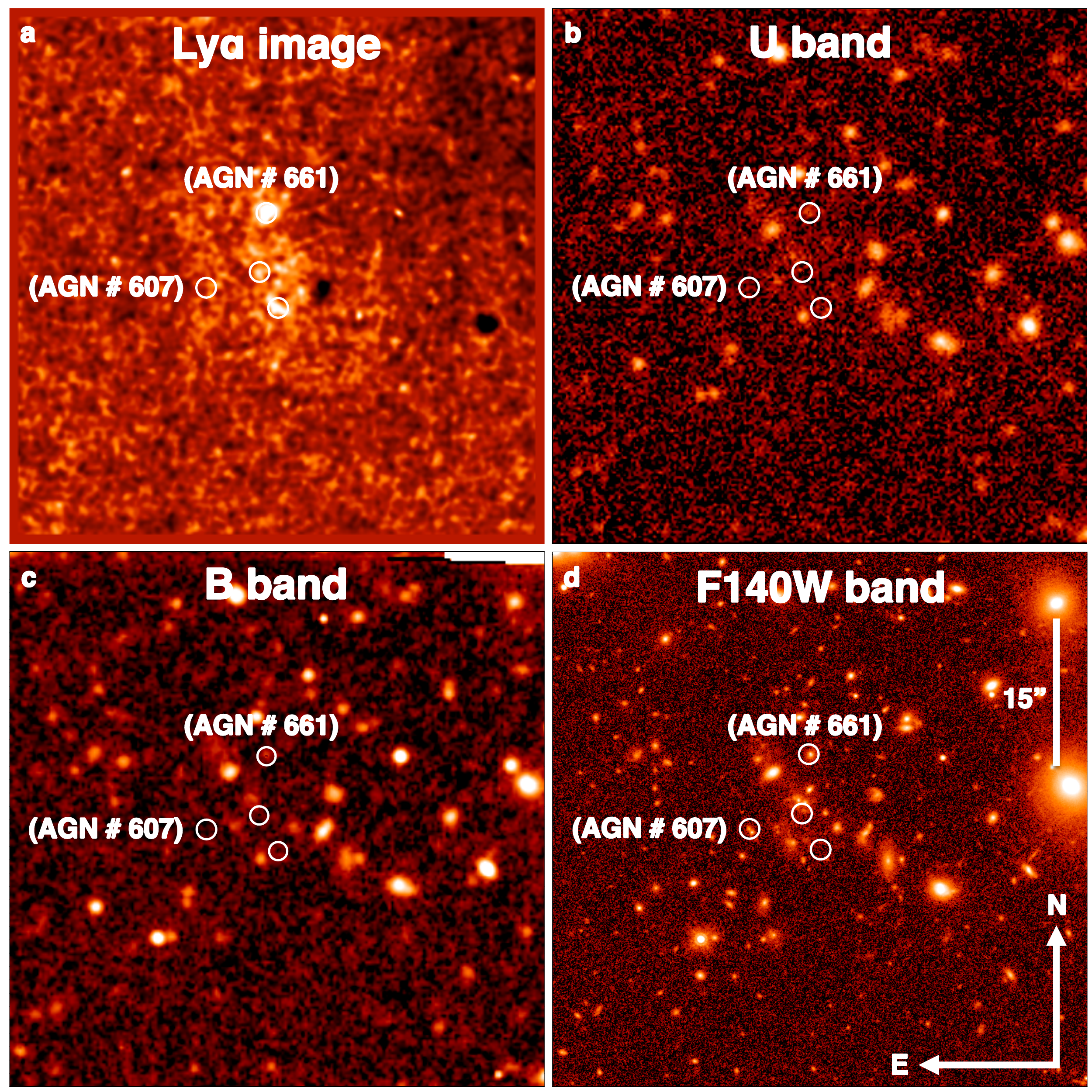}
  \caption{\textbf{Bright knots in the \lya\ nebula.} \lya\ emission
    line map smoothed over a 1" area (panel \textbf{a}) and \textit{U}-band
    (panel \textbf{b}), \textit{B}-band (panel \textbf{c}), and F140W band
    (panel \textbf{d}) images of CL~J1449+0856. Individual \lya\ emitters detected at
    $\sim5\sigma$ are marked (white circles). The position of
      the AGN \# 607 is reported for clarity, but this source is not identified as
    a \lya\ emitter. For reference,
      15'' correspond to $\sim125$~kpc at $z=1.99$.}
  \label{fig:peaks}
\end{figure}

We observed CL J1449+0856 (Figure \ref{fig:rgb}) for 3.5~h with the narrow-band filter NB3640 installed
in the blue arm of the Keck/LRIS camera on March 27, 2014, reaching a magnitude limit of
$27.8$ (5$\sigma$) in a $r = 0.6$'' circular aperture. The
  average seeing during the observation was $0.79$'' (full width half maximum). We processed the images in a standard way with the
publicly available LRIS pipeline\footnote{\texttt{http://www.astro.caltech.edu/$\sim$dperley/programs/lpipe.html}}. In
particular, we modeled and
subtracted a super-sky image obtained as the
clipped median of all the widely dithered, processed frames. We co-added
individual frames weighting them by
measured seeing and transparency variations during the observing
night. We then combined the final narrow-band image with an
aligned \textit{U}-band frame from VLT/FORS2 (5$\sigma$ limiting magnitude of
$28.1$, S13) using the formalism presented in
\cite{bunker_1995} to obtain
a \lya\ emission map (Figure \ref{fig:images} and
\ref{fig:appendix} in Appendix). Color
corrections are negligible, given the optimal overlap of the central
effective wavelengths of the narrow- and broad-band filters (3640, 3607
\AA\ respectively). We checked the absolute flux calibration against Sloan Digital
Sky Survey data, finding an agreement within $0.01$ magnitude. 
We selected individual \lya\ absorbers and emitters by running
SExtractor in dual image mode on a $\chi^2$ detection image and on
narrow- and broad-band images. We built the $\chi^2$
detection image averaging the \textit{U} and NB3640 frames weighting
by their signal-to-noise ratio squared. Besides an obscured AGN
(\#661 in G13, see \ref{sec:agn} below for further details), we
detected only two individual bright peaks in the \lya\ emission map of
the cluster core
($\sim5\sigma$) both through a classical
aperture photometry approach and a wavelet analysis
(Figure \ref{fig:peaks}). However, they are
not associated with known cluster members within a 1'' radius in the
adjacent \textit{U} and \textit{B} bands, nor in the deeper,
but redder HST/F140W band or in the X-ray bands, suggesting
  that these peaks are not associated with SFGs in the cluster
  core. The uncertainty on the position of $5\sigma$
  peaks of \lya\ emission is 0.07". The bright knots may just
be the densest regions of the extended \lya\ nebula and the
granularity (Figure \ref{fig:images}, panel c) could suggest the presence of
gas substructures \citep{cantalupo_2014} or shock fronts currently beyond our detection threshold. 
To further confirm the detection of \lya\ photons on
large scales, we performed a wavelet analysis with an iterative
multi-resolution thresholding and a Gaussian noise
model\footnote{\texttt{http://www.cosmostat.org/software/isap/}} \citep{starck_2010}. The basic concept underlying
wavelet decomposition is to split an image into a set of spatial
frequencies, each one including the signal from sources with power on
that scale. The original image is exactly recovered by adding all the
``slices''. The advantage of
this technique is to reduce (or remove) the impact of
small--scale objects when looking for large--scale structures and its
efficacy for detecting \lya\ nebulae has already been shown
\citep{prescott_2012, prescott_2013}. We used this technique for the
purpose of visualization (Figure \ref{fig:rgb}, \ref{fig:images}, and
\ref{fig:appendix} in Appendix) and to cross-check the results
from a classical aperture photometry approach. After subtraction of
the contribution from the point-like, obscured AGN ($<8$\% of the total emission), we measure a total flux of
$(8.1\pm0.8)\times10^{-16}$ \esc\ in a $\sim 140$ arcsec$^2$
polygonal aperture enclosing the whole nebula, fully
consistent with the results provided by the wavelet analysis. The residual \lya\
flux surrounding the AGN in the wavelet image is extended on
scales larger than the point spread function (PSF, with a full width
half maximum of $0.79$'') and contributes to the
luminosity of the nebula. We also retained the flux from the other
individual bright peaks since no counterparts are
detected in any other band. We estimated the $1\sigma$ uncertainty by
rescaling the photometric error measured in circular apertures with a
diameter of 1.5$\times$ the PSF ($d=1.2$''). The total flux
corresponds to an observed luminosity of
$L_{\mathrm{Ly\alpha}}=(2.3\pm0.2)\times10^{43}$ \es.\\
The morphology of the \lya\ nebula is elongated from AGN \#661
  towards the center of the cluster, suggesting a physical connection
  (Section \ref{sec:agn}). The asymmetric shape and the
    mis-centered location of the AGN is observed in several other
    nebulae at high redshift \citep[i.e.,][]{borisova_2016} and it
    might simply reflect the AGN illumination cone and
    the gas distribution in the cluster, which naturally
    concentrates towards the bottom of the potential well. In fact, the peaks of the \lya\ luminosity and the extended X-ray
emission traced by XMM-Newton and Chandra (Section \ref{sec:chandra})
are spatially coincident in projection, and so is the peak of
the stellar mass density distribution (Figure \ref{fig:xray}), implying
that the nebula effectively sits in the cluster core. Note that the peaks
mapped by XMM-Newton and Chandra are consistent within the positional
uncertainties (16'' and 4'' respectively). In Figure \ref{fig:profile}
we show the radial profile of the \lya\ surface brightness and the
projected stellar mass density. For both profiles we fixed the same
center at the peak of the projected stellar mass density
distribution. Moreover, we merged the measurements at the two
farthest positions from the cluster center to increase the
signal, and we subtracted the contribution of AGN \#661. As opposed to
the stellar component that traces the cluster potential well, the \lya\ 
surface brightness profile appears flat over the whole extension of the
nebula. A drop is expected to occur at some radius, but Figure
\ref{fig:profile} suggests that this happens at larger scales than for
the stellar component.

\subsection{Extended continuum emission}
\label{sec:continuum}
We measured the continuum emission associated with the \lya\ nebula
from a pure ``continuum emission map'' \citep{bunker_1995}
and both Subaru/Suprime-Cam \textit{B} (G11) and Keck/LRIS \textit{V}
band imaging. We do not expect strong emission lines from sources at
$z=2$ to fall in the observed \textit{B} and \textit{V} bands. These
frames are respectively $2.5\times$
and $3.3\times$ deeper than the continuum image and provide a better
constraint on the \lya\ equivalent width of the nebula (EW(\lya)). 
In unobscured SFGs, the flux density $F_{\nu}$ is roughly constant
at wavelengths bluer than $2000$~\AA, and thus possible color biases
in the evaluation of the EW(\lya) using \textit{B} and \textit{V} bands
continuum are limited. We measured the continuum emission only where we detected
the extended \lya\ emission at more than $5\sigma$
significance (Figure \ref{fig:appendix} in Appendix, panel d). Evident \textit{B} and \textit{V} band
sources were masked so as not to contaminate the diffuse emission. In
none of the frames we individually detected a significant integrated continuum emission.
Assuming a constant $F_{\nu}$ and combining the three bands, we
estimated an average continuum emission of $(3.38\pm 0.95)\times
10^{-19}$ \esc and a corresponding \lya\ equivalent width of
EW(\lya)~$ = 271^{+107}_{-60}$~\AA, compatible with the $2\sigma$ lower limit we derived from the sole
continuum image (EW(\lya) $> 192$ \AA). We note here that the
$3\sigma$ detection is formally reached only by including the
\textit{V}-band, which could contain residual contaminating emission
from red passive galaxies. Thus, it would be appropriate to regard the
quoted EW measurement as a lower limit.

\subsection{Chandra X-ray observations}
\label{sec:chandra}

\begin{figure*}
  \centering
  \includegraphics[width=\textwidth]{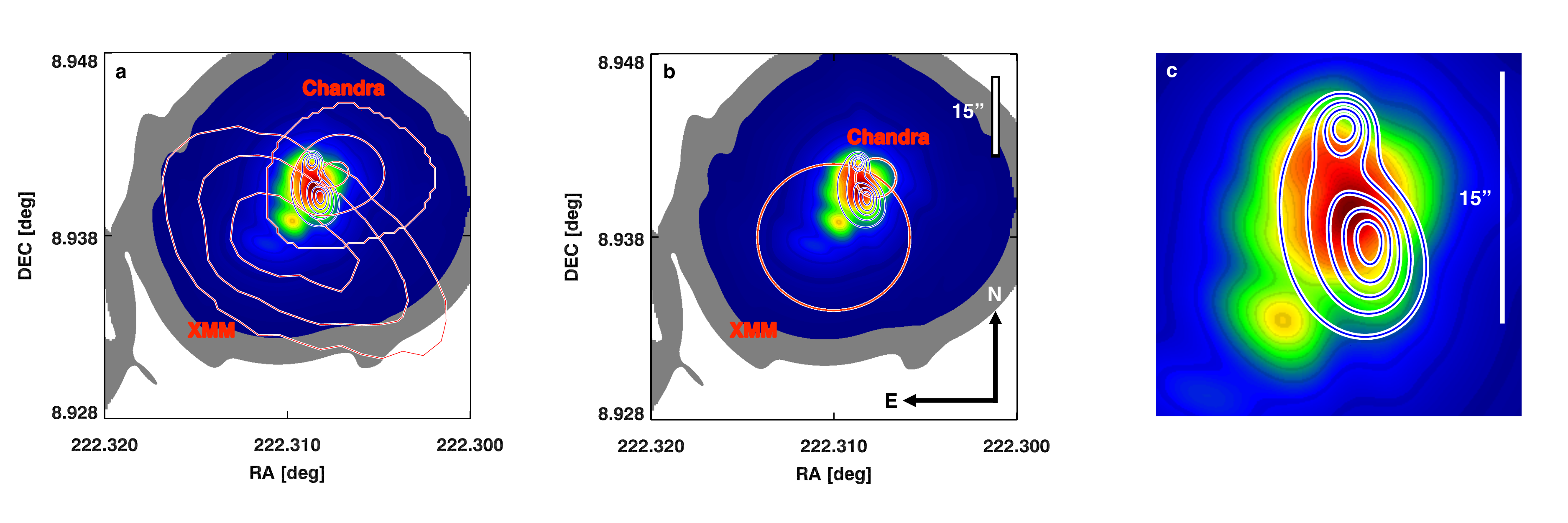}
  \caption{\textbf{Spatial distributions of stellar mass density, 
      \lya\ surface brightness, and X-ray extended emission.} 
   Stellar density maps are derived from a mass complete sample
   of cluster members and candidates with $M_{\star} \geq
   10^{10.4}$ \msun\ (S13, background colored image in panels \textbf{a}, \textbf{b}, and
   \textbf{c}). The prominent
   stellar mass density peak represents the cluster core region (red
   area). \lya\ nebula $\ge3\sigma$ contours 
   from wavelet reconstruction are superimposed (blue
   lines). Note that point-like sources have been subtracted
     before tracing the contours. Extended X-ray
   contours from XMM-Newton and Chandra
   observations (red lines) are displayed in panel \textbf{a}. The positional uncertainties
   of the peak of the X-ray extended emission from both sets of
   observations are shown in panel \textbf{b} (red
   circles). A zoom on the central region is shown in panel
   \textbf{c}. For reference, 15'' correspond to $\sim125$~kpc at $z=1.99$.}
  \label{fig:xray}
\end{figure*}

CL J1449+0856 has been imaged both with XMM-Newton (80~ks, G11,
\citealt{brusa_2005}) and Chandra (94~ks, \citealt{campisi_2009}). Details of the
XMM-Newton detection have already been reported in G11. However, that
analysis suffered from a large uncertainty on the localization of the
cluster center. Based on the statistical analysis of galaxy groups in
COSMOS \citep{george_2011}, the
difference between the most massive members and the X-ray
peak positions is typically $15^{\prime\prime}$ (Figure \ref{fig:xray},
panel b). The
measured distance between the core of the cluster and the XMM position is
$9^{\prime\prime}$.\\
Archival ACIS-I Chandra observations of the field
consist of a mosaic of three partially overlapping pointings of
$\approx$30 ks each, covering a total area of $\approx$500 arcmin$^{2}$
at different depths. These three observations (5032, 5033, and 5034)
were performed in June 2004 by the Advanced CCD Imaging Spectrometer (ACIS)
with the I0 CCD at the aimpoint and all 
ACIS-I CCDs in use. Faint mode was used for the event telemetry, and
ASCA grade 0, 2, 3, 4, and 6 events were used in the analysis (full
details are reported in \citealp{campisi_2009}). In Cycle~16 we
followed-up the field with the ACIS-S camera (aimpoint at CCD=7) for a
nominal exposure of 94.81 ks in very faint mode. This new Chandra observation has a higher spatial
resolution because pointed at the location of the diffuse emission and, thus,
improves the localization of the cluster core and the association between the extended X-ray source and
the optical/near-IR counterpart. For both ACIS-I and ACIS-S data,
reprocessing was carried out using CIAO version 4.6 and adopting the
latest relevant calibration products. From a wavelet reconstruction of the
ACIS-S image, we detected a $>4\sigma$ extended feature co-aligned with
the core (Figure \ref{fig:xray},
panel a). The X-ray source is centered on coordinates 14:49:13.670, +8:56:28.25
with a $1\sigma$ uncertainty on the position of $4^{\prime\prime}$ (Figure \ref{fig:xray},
panel b) and
a distance to the cluster core of $5^{\prime\prime}$. We measured the
extended source flux in the area where the significance of the wavelet map was higher than 2$\sigma$. 
We derived ACIS-S and ACIS-I counts independently, using the same extraction region.
Within a 10$^{\prime\prime}$ aperture, the net (i.e., background-subtracted) number of counts from
the extended source in ACIS-S is $11.0\pm5.3$ (94 ks exposure) in the $0.5-2$
keV band, corresponding to an aperture flux of $(8.5 \pm 3.0) \times 10^{-16}$
\esc. The ACIS-I counts and aperture flux are $5.2\pm2.5$ and $1.2
\times 10^{-15}$ \esc\ respectively (49 ks exposure).
The average flux of the source is therefore $(1.0\pm0.4) \times
10^{-15}$ \esc. This corresponds to an observed total X-ray luminosity
of $L_{\rm X} = (9\pm3)\times10^{43}$ \es\ in the $0.1-2.4$ keV
rest-frame band within $R_{500}$, defined as the radius enclosing a
mean overdensity $500\times$ larger than the critical density of the Universe.
We do not detect bright
radio sources close to the cluster core in deep Jansky Very Large
Array observations at
3~GHz down to $2.7\,\mu$Jy (rms), except for two galaxies with a
$\sim30$ $\mu$Jy continuum emission, fully consistent
with pure star formation activity seen at ultra-violet and infra-red
wavelengths. Thus, inverse Compton scattering off extended
radio-galaxy jets is not likely to be the origin
of the extended X-ray emission as in potentially similar cases
\citep[i.e.,][]{miley_2006}.

\subsection{Halo mass and gas temperature}
\label{sec:halo}
Scaling the observed total X-ray luminosity within $R_{500}$ \citep{leauthaud_2010}, we
estimated a halo virial mass of $M_{\mathrm{halo}} = (5-7) \times
10^{13}$ \msun\ and a virial radius of $R_{\mathrm{vir}} = 0.5\pm
0.1$ Mpc, in agreement with previous determinations (G11, G13). This
estimate is consistent with that expected from a total
stellar mass enclosed in cluster members of $2\times10^{12}$ \msun, in particular in six
massive and passive galaxies in the core (S13, $M_{\mathrm{halo}} =
4-7\times10^{13}$ \msun, including the latest calibration by
\citealp{vanderburg_2014}). We independently evaluated
$M_{\mathrm{halo}}$ from the velocity dispersion derived from
\textit{HST}/WFC3 and Subaru/MOIRCS spectroscopy
\citep[G13,][]{valentino_2015}. After excluding obvious
interlopers at redshift $z<1.95$ and $z>2.05$, we estimated the
systemic redshift and the velocity dispersion fixing the reduced
$\chi_{\rm red}^2 = (\sum_{i=1}^{N} (z_{\rm i} - z_{\rm sys})^2 /
(\sigma_{z_{\rm i}}^2 + \sigma_{\rm disp}^2)) / \mathrm{dof} = 1$, applying a clipping at
$3\sigma$, and iterating until convergence. This procedure allows us to fully take
into account the uncertainties on spectroscopic redshifts.
We then estimated the uncertainties as the 15.87 -
84.13 percentile ranges of the distribution of 15,000 bootstrap
simulations. We obtain $z_{\rm sys} =
1.995^{+0.003}_{-0.004}$ and $\sigma_{\rm vel} =
(830\pm 230)$ \kms. We find consistent results modeling a Gaussian
curve on the galaxy redshift distribution (Figure
\ref{fig:dispersion}). Assuming virialization, we find
a $1\sigma$ lower limit on the virial mass of $M_{\mathrm{halo}} \gtrsim
4\times10^{13}$ \msun\ obtained adopting the $1\sigma$ lower limit
on $\sigma_{\rm vel}$. Then, we calculated a total intracluster mass in the hot phase of
$M_{\mathrm{ICM}} \approx 0.08 \times M_{\mathrm{halo}} \approx 5 \times
10^{12}$ \msun\ \citep{renzini_2014}. The gas fraction may vary with
redshift, but even considering a value close to the universal baryon
fraction, the main result of this work would not change. Assuming
spherical geometry for the halo and a mean molecular weight of $\mu=0.6$, the
average particle density is $(8\pm2)\times 10^{-4}$ cm$^{-3}$ within the virial radius.
Finally, we estimated a temperature of $2.1$~keV from the $L_{\mathrm{X}}-T$
relation \citep{finoguenov_2007} and an absorbing column density of
$N_{\mathrm{H}}$=$2\times10^{20}$ cm$^{-2}$. We stress here that the
current X-ray dataset allows only for an estimate of the integrated
X-ray luminosity $L_{\rm X}$. We do not have in-hand the spatial profiles of X-ray
derived quantities such as the temperature, entropy, density, or the
metallicity of the hot ICM. In order to estimate these physical
quantities, we rely on the extrapolation of well established relations
at low and moderate redshift ($z<1$).

\section{Physics of the \lya\ nebula}
\label{sec:physics}
In this section we study the physics of the \lya\
nebula. First, we estimate the mass and electron density from its
luminosity and size. We then explore the possible powering
mechanisms and conclude that the most plausible source of ionizing
photons are AGN embedded in the nebula, with a possible
  contribution from dissipation of the mechanical energy due to galaxy
outflows. Finally, we discuss the
typical timescales regulating the evolution of the nebula. We find that,
barring an observational coincidence, in our favored scenario
  the nebula is constantly
replenished with cold gas to survive evaporation due to the surrounding
hot X-ray plasma. 

\subsection{Mass and density}
\label{sec:lya_mass}
Assuming photoionization, we can estimate the mass $M_{\mathrm{Ly\alpha}}$ and the electron
density $n_{\rm e}$ of the ionized gas from the \lya\ luminosity
\citep{mccarthy_1990, dey_2005}:
\begin{equation}
\label{eq:mass}
M_{\mathrm{Ly\alpha}} = 1.25 \, m_{\mathrm{p}} \, n_{\mathrm{e}} \, f
\, V = (1 - 10)\times 10^{9}\,\mathrm{M_\odot}
\end{equation}
where $m_{\mathrm{p}}$ is the proton mass, $f$ the volume filling factor, and $V$ the volume of
the nebula. For the sake of simplicity, we assumed a spherical geometry for the nebula with a radius
$R_{\mathrm{neb}}=46$ kpc, the average value of the long and short axes measured in
the wavelet reconstructed image. The choice of the shape does not
affect the final result of this work, i.e., adopting a cylindrical
symmetry the volume changes by $\approx 10$\%. We assumed
$f=10^{-3}-10^{-5}$ as detailed in next section. The
electron density is derived from the \lya\ luminosity estimate through:
\begin{equation}
L_{\mathrm{Ly\alpha}} =
\frac{j_{\mathrm{Ly\alpha}}}{j_{\mathrm{H\beta}}}\,
\alpha_{\mathrm{H\beta}}^{\mathrm{eff}} \, h\nu_{\mathrm{H\beta}}
\, n_{\mathrm{e}}\,n_{\mathrm{p}}\,f\,V \rightarrow  n_{\mathrm{e}} = 0.9 - 9\,\mathrm{cm^{-3}}
\end{equation}  
where $j_{\mathrm{Ly\alpha}}$ and $j_{\mathrm{H\beta}}$ are the
emission coefficients for \lya\ and \hb,
$\alpha_{\mathrm{H\beta}}^{\mathrm{eff}}$ is the effective
recombination coefficient for \hb, $h\nu_{\mathrm{H\beta}}$ is the
energy of an \hb\ photon, and $n_{\mathrm{p}}$ the proton
number density ($n_{\mathrm{e}}\approx 1.2\,n_{\mathrm{p}}$ accounting
for doubly ionized helium). The range of $n_{\mathrm{e}}$  values
corresponds to $f=10^{-3}-10^{-5}$, assuming case B recombination
\citep{osterbrock_2006} and
$T=10^4$ K. We notice that the gas appears marginally optically thick
to ionizing radiation, given the column density of neutral hydrogen
averaged over the projected area of the nebula of $\langle
N_{\mathrm{HI}} \rangle \thickapprox 10^{17.2}$~cm$^{-2}$
\citep[Eq. 11]{hennawi_2013}. Moreover, $n_{\mathrm{e}}\propto \left( \sqrt{f} \right) ^{-1}$ and
$M_{\mathrm{Ly\alpha}}\propto \sqrt{f}$, reducing the 2 orders of
magnitude range of uncertainty that we allowed for $f$. Finally,
$M_{\mathrm{Ly\alpha}}$ might be a lower limit for the total
mass of cold gas reservoirs in the cluster if AGN are the powering
sources (see Section \ref{sec:powering}), as beamed
emission may illuminate only a portion of the gas. In addition, the
true \lya\ luminosity may be higher than reported due to
dust and neutral hydrogen absorption.
\subsection{Volume filling factor}
\label{sec:volume_filling_factor}
The mass and density of the nebula depend on the volume
  filling factor $f$, which is not directly constrained by our
  observations. However, it is reasonable to assume pressure
  equilibrium between the ionized gas and the hot X-ray ICM, allowing
  us to put an upper limit on the possible
  values of $f$. To estimate the pressure exerted by the hot ICM, we
  assumed the universal pressure profile 
  of galaxy clusters \citep{arnaud_2010}, properly rescaled in
 mass and redshift, as representative for CL~J1449+0856.
  Then, dividing the pressure by
  $\sim10^4$~K, the typical temperature of the \lya\ gas, we obtained
  the radial density profile of a medium in pressure equilibrium
  with the X-ray emitting plasma. The range of possible values of
  $n_{\rm e}$ over the radial extension of the nebula is $n_{\rm
    e}\sim1-10$~cm$^{-3}$, corresponding to $f\sim10^{-3}-10^{-5}$, a pressure of
  $p\sim10^4-10^5$~K~cm$^{-3}$, and masses of ionized gas of
  $M_{\mathrm{Ly\alpha}}\sim(1-10)\times10^9$ \msun. Absent an
  observed X-ray profile, this is an order of magnitude calculation,
  given that the pressure profile in low mass systems might be
different and, notably, flatter than in clusters \citep{lebrun_2015},
leaving the door open for larger values of $f$ and lower densities. However,
in addition to pressure equilibrium, higher values of
$f$ ($\sim0.01-1$) are disfavored by a simple 
argument based on gravitational stability: if larger and more massive
clouds were in place, they would be Jeans-unstable and
form new stars, a scenario disfavored by the
observed high value of EW(\lya) (Section \ref{sec:sfr}). On the contrary, solutions
with $f\lesssim10^{-3}$ are gravitationally stable, considering the
simplified case of auto-gravitating spheres of gas at $10^4$~K stably ionized.\\
 On the other hand, much smaller values of $f$ are not easily
 maintained for long timescales. As recognized in classical works
 \citep{fabian_1987, crawford_1989}, lower volume
filling factors and higher densities would imply clouds 
dissipating by thermal expansion on short timescales ($10^5$
yr), with consequent difficulties to explain the size of the
nebula and its lifetime.

\subsection{Powering mechanism and origin of the gas}
\label{sec:powering}
We consider five different physical scenarios to explain the
  extended \lya\ emission: hard ionizing spectra of AGN impacting
  gas reservoirs in the cluster core, the continuous formation of
  young massive stars, 
  cooling of dense cosmological cold flows penetrating into the dark
  matter halo, cooling of plasma from the X-ray
  phase, and dissipation of the mechanical energy from galaxy outflows
  in the core. Eventually, we point the AGN radiation field as the most
  plausible powering source of the \lya\ nebula, with a potential
  contribution from shocks induced by galaxy outflows.

\subsubsection{AGN in the cluster core}
\label{sec:agn}
\begin{figure}
  \includegraphics[width=\columnwidth]{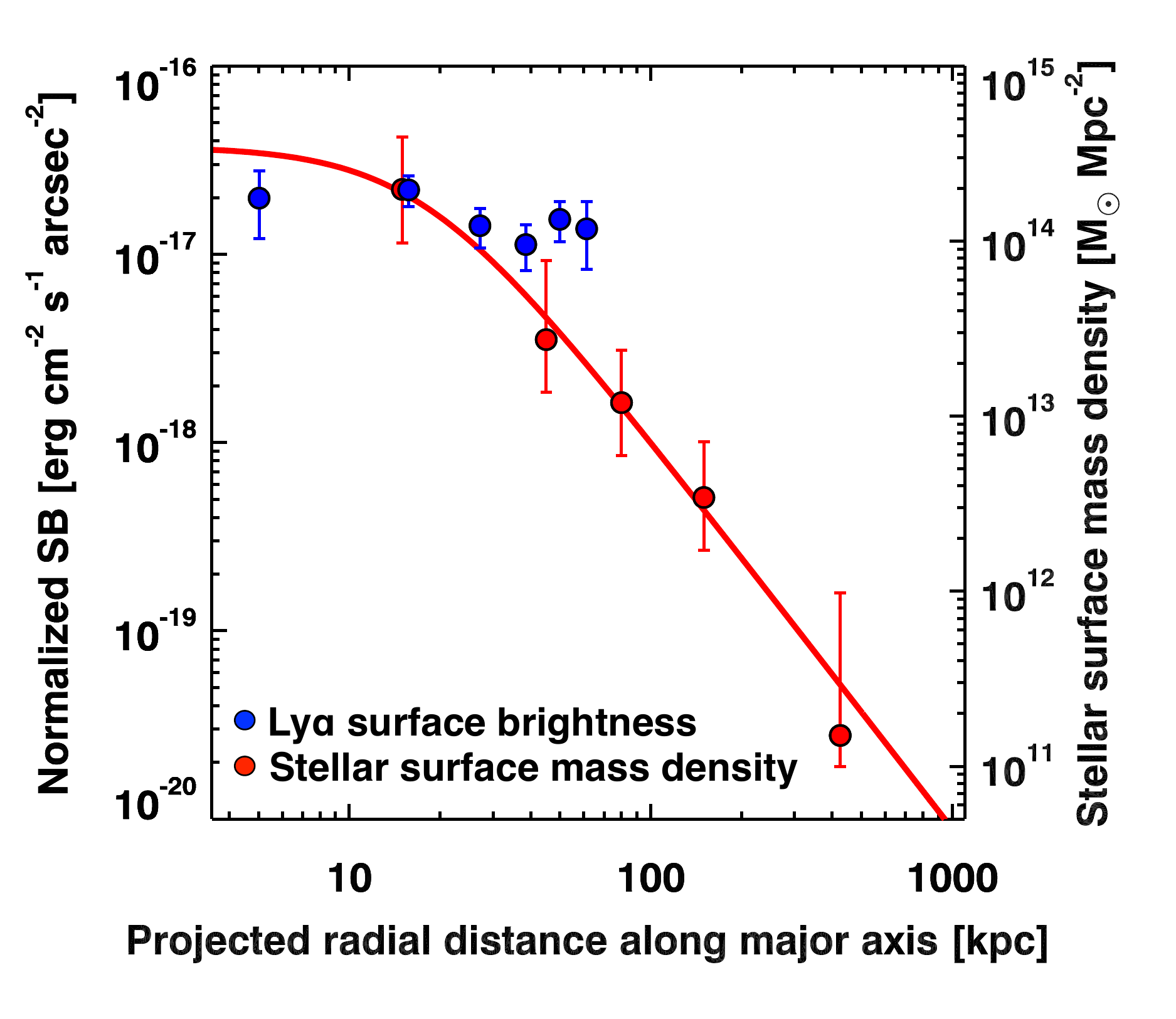}
  \caption{\textbf{\lya\ nebula and projected stellar mass density profiles.} We show the stellar mass
    density (red circles) and normalized \lya\ surface
    brightness (blue circles) radial profiles. The center of
      the profiles are spatially coincident and fixed at the position
      of the barycenter of stellar mass (S13). For the \lya\
        surface brightness profile we merged the measurements at the two
farthest positions from the cluster center to reach the formal
detection threshold. Moreover, the \lya\ flux of AGN \#661 has been
      subtracted in the corresponding bin. For the stellar mass density
    profile, error bars include both the Poisson error and the
    uncertainties in membership determination (S13). For the surface
    brightness profile, error bars represent the $1\sigma$ uncertainties on flux
    measurements. The best fit of the stellar mass profile is a
    classical beta-model (S13).}
\label{fig:profile}
\end{figure}
Two spectroscopically confirmed X-ray AGN in the cluster core (\#607, 661 in
G13) are suitable candidates for ionizing the nebula. The depth of the
new Chandra observation, coupled with an optimal on-axis alignment, allowed
us to perform a basic X-ray spectral analysis despite the limited
photon statistics (34 and 20 net counts in the observed $0.5-7$~keV
band for sources \#607 and 661, respectively). 
Source \#607 is characterized by a power-law spectrum with photon index 
$\Gamma=2.0\pm{0.6}$; the observed $2-10$~keV flux is 
$1.7^{+1.1}_{-0.6}\times10^{-15}$ \esc, corresponding to a rest-frame $2-10$~keV 
luminosity of $5.2^{+3.4}_{-1.8}\times10^{43}$ \es, typical of a luminous Seyfert galaxy. 
The X-ray spectrum of source \#661, the point-like \lya\ emitter
(Figure \ref{fig:images}), is flat: fitting the data with a power-law model provides 
$\Gamma=-0.7^{+0.8}_{-0.9}$, highly indicative of strong obscuration. We then included an 
absorption component and fixed the photon index to 1.8, as expected for the intrinsic AGN 
emission \citep[e.g.,][]{piconcelli_2005}. This model results in a column density of 
$N_{\rm H}=9.3^{+5.6}_{-4.0}\times10^{23}$~cm$^{-2}$, i.e., consistent
with marginal Compton-thick 
absorption ($1.5\times10^{24}$~cm$^{-2}$). The tentative detection of an iron K$\alpha$ 
emission line at 6.4~keV (with equivalent width of $\approx2.4$~keV rest frame), if confirmed, 
would further support the heavily obscured nature of source \#661. The derived $2-10$~keV flux is 
$(7.4\pm2.2)\times10^{-15}$ \esc, corresponding to a rest-frame
luminosity of $L_{\rm 2-10\,keV}=2.9^{+0.6}_{-0.5}\times10^{44}$ \es, placing
source \#661 in the quasar regime. We do not detect any bright
counterpart in deep Jansky Very Large Array observations at
3~GHz down to $2.7\,\mu$Jy (rms), and we thus classify source \#661 as radio-quiet. From
aperture photometry, we estimated a
\lya\ flux of $(6.7\pm
0.7)\times 10^{-17}$
\esc, corresponding to a luminosity of $(1.9\pm0.2) \times 10^{42}$
\es. The spectral energy distribution (SED) of \#661 is shown in Figure \ref{fig:sed}. From SED modeling, 
which benefits from near-, mid- and far-IR observations from Spitzer and Herschel, 
we estimated a bolometric luminosity for the AGN of
$(2.7\pm1.5)\times10^{45}$ \es. A similar value ($3.2\pm0.6\times10^{45}$ \es) is derived using the
observed \oiii$\lambda5007$ \AA\ luminosity obtained from recent
Subaru/MOIRCS spectroscopy of the galaxy \citep{valentino_2015},
converted into a bolometric luminosity as
$L_{\mathrm{bol}}/L_{\mathrm{[OIII]}} = 3500$ \citep{heckman_2004}. Assuming the luminosity-dependent bolometric correction as in 
\cite{lusso_2012}, we predict an intrinsic $2-10$~keV luminosity for source 
\#661 of $1.6^{+1.6}_{-0.5}\times10^{44}$ \es. This value is consistent, within the 
uncertainties due to the adopted relations and measurements, with that derived from the X-ray 
spectral analysis reported above.\\
Furthermore, we normalized the ``radio-quiet AGN'' template by
  \cite{elvis_1994} to match the estimated $L_{\rm bol}$. We then
  integrated over wavelengths bluer than the Lyman continuum limit
  to obtain the ionizing photon flux $\phi$ from
both sources. We obtained $\phi\sim1.3\times10^{55}$ and
$\phi\sim7.3\times10^{54}$ photons s$^{-1}$ for source \#661 and
\#607, respectively. Taking into account the distance between the AGN and the
peak of diffuse \lya\ emission, a conical illumination pattern of
the neutral gas, and a covering factor of the ionized gas $f_{\rm
  C}\sim1$ consistent with the observations (Section
\ref{sec:timescales}), we estimate that $(6.5-15.3)$\% and $(14.5-49.2)$\% of
ionizing photons from \#661 and \#607 reach and ionize the
gas. The number of ionizing photons necessary
to explain the observed \lya\ luminosity is:
\begin{equation}
\phi = \frac{L_{\mathrm Ly\alpha}}{h\,\nu_{\mathrm{Ly\alpha}}}\,\frac{1}{\xi_{\mathrm Ly\alpha}} \approx 1.8\times10^{54}\,\mathrm{photons\,s}^{-1}
\end{equation}   
where $\xi_{\rm Ly\alpha} = 0.68$ is the fraction of ionizing photons
converted in \lya\ \citep{spitzer_1978}. Thus, the AGN are likely capable of
producing a sufficient amount of ionizing radiation to power the gas
emission, even if $f_{\rm C}$ were a factor of several times
  smaller. We note that the flat \lya\ surface brightness distribution
  in Figure \ref{fig:profile} is not \textit{a priori} in
contradiction with powering from the AGN. The geometry of the system,
the absorbing torus around the AGN, and the distribution of the cold
clouds impact the observed profile: the flatness might just reflect
covering factors close to unity. In fact, for volume filling factors
$f=10^{-3}-10^{-5}$ and a covering factor $f_{\rm C}\sim1$, energetic
photons from the AGN may ionize gas at large distances. Finally, we note that resonant
 pure scattering of \lya\ photons from \#661 can hardly contribute to the
 diffuse luminosity farther than $\sim10$ kpc - less than 10\% of the
 whole extension of the nebula -, as detailed radiative transfer
 modeling shows \citep{cantalupo_2005, dijkstra_2006}. 

\subsubsection{Young massive stars}
\label{sec:sfr}
\begin{figure}
  \centering
  \includegraphics[width=\columnwidth]{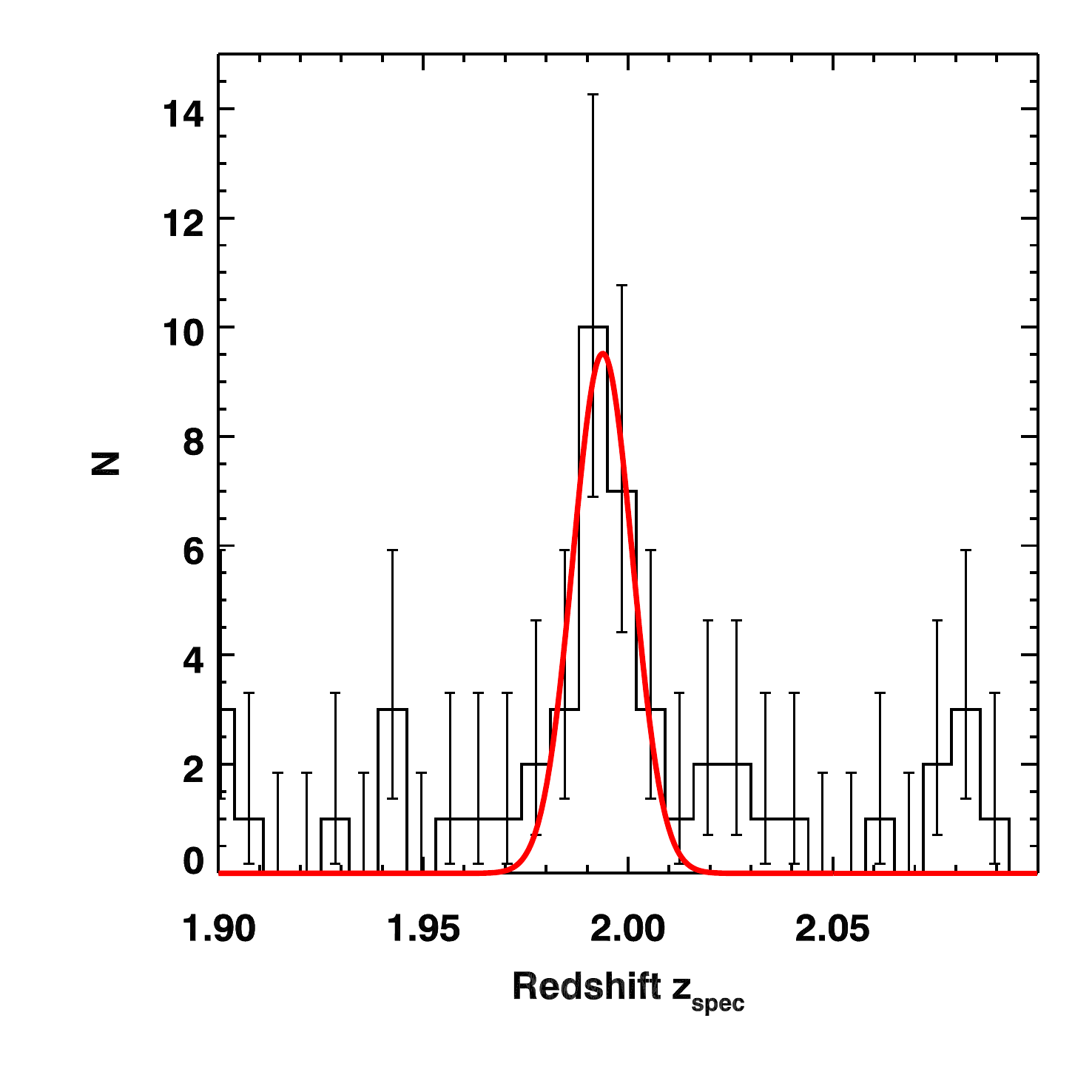}
  \caption{\textbf{Cluster redshift distribution.} The black histogram shows
  the redshift distribution of galaxies in CL~J1449+0856. The red
  curve is the best Gaussian model fitting the curve. The
  uncertainties represent the formal $68.3$\% Poisson confidence
  interval.}
 \label{fig:dispersion}
\end{figure}
Ongoing and continuous formation of young stars spread over the nebula
might be a possible alternative ionizing source
\citep{miley_2006}. The total star formation rate (SFR) inferred from the \lya\ luminosity
is $21\pm2$ \myr\ \citep{kennicutt_1998}, assuming an intrinsic ratio of 
$L_{\mathrm{Ly}\alpha}$/$L_{\mathrm{H}\alpha} = 8.7$ \citep[Case B
recombination,][]{osterbrock_2006}. This estimate should be regarded
as a strong lower limit on the total SFR since we do not correct
$L_{\mathrm{Ly}\alpha}$ for dust obscuration and scattering from neutral
hydrogen. However, both truly diffuse star formation and the emergence
of undetected galaxies populating the low-mass end of the mass
function and contributing to the diffuse emission \citep{zibetti_2005}
are disfavored: the high value of EW(\lya)$=271^{+107}_{-60}$ \AA\ implies
ages too young to be reasonably observable (Section
\ref{sec:continuum} and Figure 7 in \citealt{schaerer_2003}). Assuming a continuous star formation history, we should observe
stars younger than $\lesssim3$ Myr distributed over a 100-kpc scale, much
larger than the typical super-star cluster size. For comparison,
EW(\lya) is $\sim100$ \AA\ for the continuous star formation regime. A
single, simultaneous star-burst
event on the same scale seems even less likely. Small effects due to
the choice of the initial mass function or metallicity do not change
these results, unless considering extreme Population III stars
\citep{schaerer_2003}. We stress here that the weak continuum
detection is formally reached only by averaging the \textit{U} frame with
redder bands, which could contain residual contaminating emission
from red passive galaxies. In addition, the \lya\ flux is not
corrected for dust absorption and scattering from neutral
hydrogen. Hence, the quoted EW measurement is reasonably a lower limit
of the true value.

\subsubsection{Cosmological cold flows}
\label{sec:coldflows}
Another viable origin for the \lya\ photons is the cooling of the dense
streams penetrating into dark matter halos currently predicted by
hydrodynamical cosmological simulations \citep{dekel_2009,
  goerdt_2010}. The current status of these models disfavors this
scenario showing that, given the halo mass of CL~ J1449+0856, these
cold flows should have stopped reaching the cluster core $\sim1$~Gyr
prior to observation, being
shock-heated to the virial temperature \citep{valentino_2015}. Nevertheless, in the cluster core we estimate a
total SFR of $\approx 1000$ \myr\ (Section \ref{sec:activity}) that must be constantly fueled by fresh
cold gas given the 0.5~Gyr gas depletion timescale typical at $z=2$
\citep{daddi_2010, tacconi_2013}. This points to an
  inconsistency with the prescriptions of present day models. Note,
  however, that the mass and redshift regimes at which cold flows
should not penetrate into the hot ICM have not been
observationally confirmed yet, and suffer from substantial scatter in
simulations \citep{dekel_2009}. In addition, there are hints that within
this scatter a cluster progenitor at $z=2$ may be crossed by dense gas
streams supporting high SFRs and
powering extended \lya\ nebulae \citep{danovich_2015}. 
For the rest of the paper, we will adopt the predictions of
current cosmological simulations, excluding a substantial contribution
to the \lya\ luminosity from cold flows. We defer a more detailed discussion to
future work.

\subsubsection{Classical cooling flows and cool-cores}
\label{sec:coolcores}
Cooling from the X-ray emitting phase to a cold $\sim10^4$ K
  temperature is known to occur at low redshift and is generally considered the origin of the
nebular filaments observed in cool-core clusters
\citep[CCs,][]{fabian_1984, heckman_1989, hatch_2007, mcdonald_2010,
  tremblay_2015}. More extreme manifestations of the
same mechanism are the classical cooling flows, though obvious
cases are not currently known in the local Universe
\citep{peterson_2006}. Even though we
cannot properly classify CL~J1449+0856 as CC or non-CC according to the
standard X-ray based definition owing to poor X-ray sensitivity at
$z=2$, we find several inconsistencies between this cluster
and the typical local CCs or classical cooling flows.
First, \textit{the ratio between the \lya\ luminosity of the nebula and
    the total X-ray luminosity of the ICM is orders of magnitude
    larger in CL~J1449+0856 than predicted for classical cooling
    flows or observed in local CCs.} In CL~J1449+0856 we find
  $L_{\mathrm{Ly\alpha}}/L_{\mathrm{X}}\sim0.3$, while
  $L_{\mathrm{Ly\alpha}}/L_{\mathrm{X}}\sim10^{-3}$ and $\sim0.5\times10^{-3}$ for classical
  stationary cooling flows \citep{cowie_1980,
  bower_2004, geach_2009} and CCs, respectively.
To compute the ratio for local CCs, we
  collected measurements of extended \ha\ luminosities from the survey
  by \cite{mcdonald_2010} and the X-ray flux observed with
  ROSAT \citep{ledlow_2003}. Assuming $L_{\rm Ly\alpha}/L_{\rm H\alpha}=8.7$ \citep[Case
  B recombination,][]{osterbrock_2006}, we derived an average
  $L_{\mathrm{Ly}\alpha}$/$L_{\rm X}$ ratio of $5\times10^{-4}$ (40\% less when including only the
  extended filaments and not the flux from the BCG) for 13 structures in the Abell
  catalog. This is a conservative upper
  limit, since our \lya\ measurement for CL~J1449+0856 is not
  corrected for reddening nor scattering. The only
  cases when $L_{\mathrm{Ly}\alpha}$/$L_{\rm X}\sim0.01$ happen in presence of strong
  radio-galaxies (i.e., Hydra A), while we exclude emission from such
  sources in CL~J1449+0856 thanks to our deep JVLA
    3~GHz maps down to $2.7\,\mu$Jy (rms). This was already recognized in the
  seminal paper by \cite{heckman_1989} where the highest
  $L_{\mathrm{Ly}\alpha}$/$L_{\rm X}$ ratios strongly correlate with the presence of
  a bright radio-galaxy in the core (Cygnus
  A, 3C~295, Perseus) and consequently show high excitation lines in the
  spectra of the nebulae. For reference, the widely studied case of the Perseus cluster \citep[i.e.,][]{fabian_1984lya,
  conselice_2001, hatch_2005, hatch_2007} shows
$L_{\mathrm{Ly\alpha}}/L_{\mathrm{X}}\lesssim5\times10^{-3}$. We
stress once more that we do not measure a proper
  observed X-ray profile for the cluster. Thus we cannot isolate the core luminosity
  \citep[better correlated with the nebular luminosities, Figure 9 and
  11 in][]{heckman_1989}, but we can only compare global properties
  (their Figure 10). Overall, the \lya\ nebula we
  discovered is hugely overluminous with respect to local analogs:
  only $\ll 1$\% of its luminosity could be explained if CL~J1449+0856 were
  the high-redshift version of a typical low-redshift CC.\\
Moreover, local nebular filaments are frequently connected with
episodes of star formation. If not in the filaments themselves - observationally there
  is not clear evidence disproving this possibility
  \citep{mcdonald_2010, odea_2010, tremblay_2015} - star formation
  should occur at least in the central galaxies, fueled by
  the gas cooled from the X-ray phase. The presence of large
  molecular gas reservoirs associated with the filaments
  \citep{salome_2011, mcnamara_2014} further supports
  this argument. In CL~J1449+0856 this is not observed: the peak of
  the extended \lya\ emission (once removed the contribution of the
  offset point-like AGN) does not overlap with any cluster member, nor to
  any evident source in all bands from \textit{U} to
  near-infrared \textit{HST}/F140W. In this sense, if cooled gas is
  flowing towards the bottom of the potential well where the peak of
  the \lya\ emission lies (Figure \ref{fig:xray}), it is not triggering star formation nor AGN
  feedback in any object.\\
Finally, as we will show later in Section \ref{sec:injection}, SFGs and AGN in
the cluster core can inject a huge amount of energy into the
surrounding medium. Considering only mechanical energy,
this quantity is $5\times$ higher than the observed X-ray luminosity
at $z=1.99$, largely enough to offset \textit{global} catastrophic cooling from
the ICM and to strongly disfavor the hypothesis of a classical
cooling flow. However, \textit{local} rapid cooling may arise at the
peak of the density distributions in the ICM, caused by onset of
thermal instabilities. This argument is at the base of modern feedback
regulated models of ICM cooling, which have proved to successfully
reproduce several properties of the local nebular filaments
\citep{gaspari_2012, sharma_2012}. 
Here we cannot directly test the
simple prescription proposed in these models based on the ratio
between the free-fall time and the timescale necessary to start the
thermal instabilities. Nevertheless, we note that feedback is likely
to play a role (Section \ref{sec:injection}), even if a circular
`on-off' auto-regulated regime might not be easily established at high
redshift, given the long gas depletion timescales in galaxies
\citep[$0.5-1$ Gyr,][]{daddi_2010, tacconi_2013} compared with the age
of the Universe.\\
We conclude that the observed \lya\ emission is not due to
cooling from the X-ray phase in the form of a classical stationary
flow. On the other hand, if moderate cooling partially contributes to
  the total \lya\ luminosity regulated by feedback, it
  generates very peculiar features not observed in local CCs.
\begin{figure}
  \centering
  \includegraphics[width=\columnwidth]{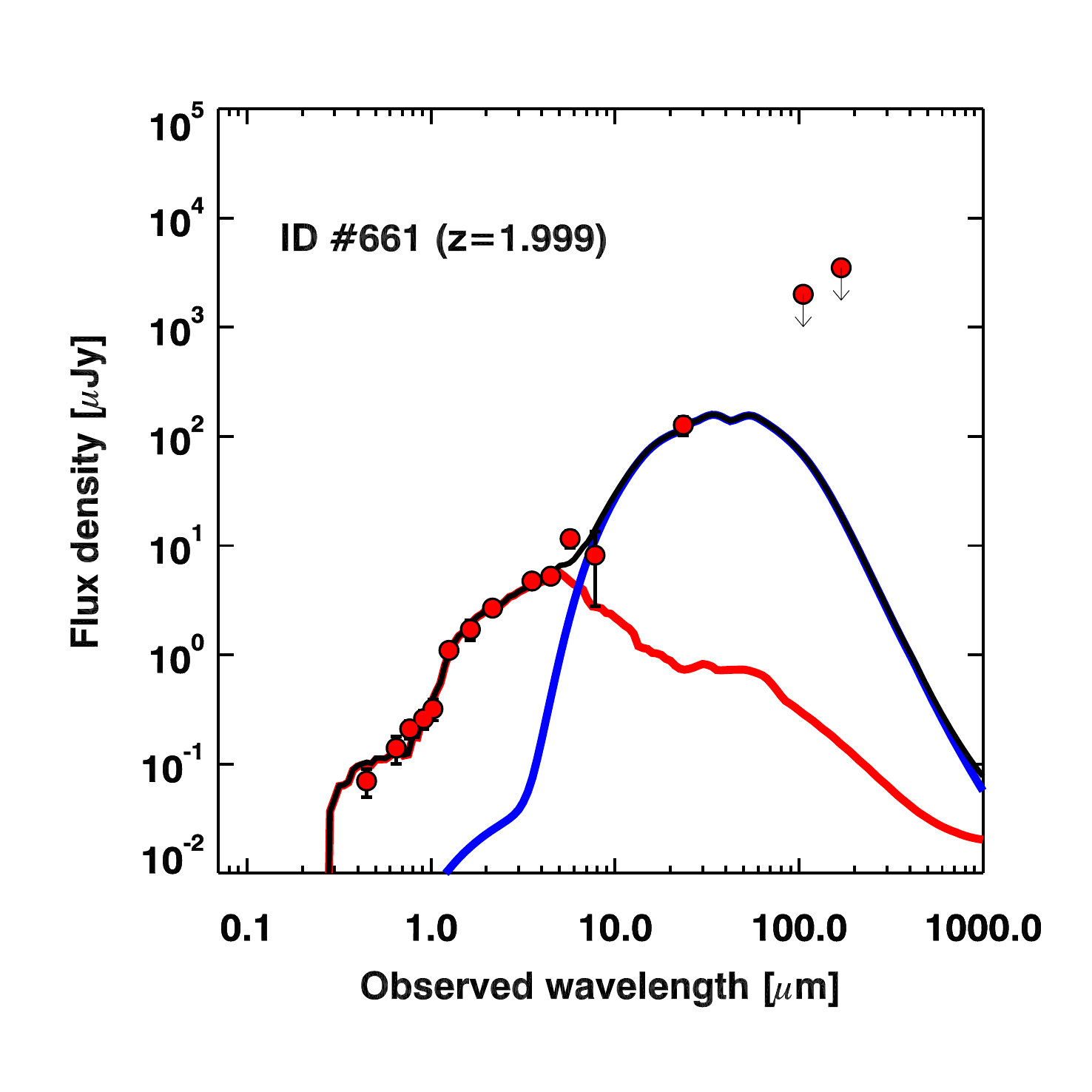}
  \caption{\textbf{SED modeling of the marginally Compton-thick
      AGN \#661.} \textit{U}-band to Herschel/SPIRE 250 $\mu$m observations are shown
    (red circles). The contributions of stars (red line) and the AGN
    (blue line) are shown independently. The full model is the sum of
    the two components (black line). The spectroscopic
      redshift is from Subaru/MOIRCS \citep{valentino_2015}.} 
  \label{fig:sed}
\end{figure}

\subsection{Shocks}
\label{sec:shocks}
\lya\ emission could be powered by shocks induced by galaxy outflows on the surrounding
pressurized ICM. We constrain the maximum
fraction of total kinetic energy injected by winds that is lost by
radiative losses simply dividing the
total power radiated through emission lines ($\approx2\times$ the observed
total \lya\ luminosity $L_{\mathrm{Ly}\alpha}$) by the instantaneous
energy injection ensuing galaxy outflows $\dot{E}_{\rm kin}$ (Section
\ref{sec:injection}). This fraction ($\sim10$\%) is presumably a strong upper
limit, considering the large number of
ionizing photons emitted by AGN and star-forming galaxies in the core,
and the low density of the ICM.
If shocks were dominating the \lya\ emission, we could not estimate the mass from Eq. \ref{eq:mass}, but rely on alternative
working hypotheses, i.e., pressure equilibrium and geometrical
assumptions. Future spectroscopic
follow-up will help to quantify the contribution of shocks to the
nebular emission, i.e., from UV lines ratios \citep{dey_2005, prescott_2009}.

\subsection{Time evolution of the \lya\ nebula}
\label{sec:timescales}
The evolution and the lifetime of the \lya\ nebula are globally
driven by cooling and heating processes, the dynamics of the gas, and
their typical associated timescales. In the following, we envisage the
time evolution of the system assuming that it is stable and
exploring different physical scenarios.
\subsubsection{Dynamics}
As mentioned in Section \ref{sec:volume_filling_factor}, a
  single massive nebula at rest at the bottom of the potential well would
rapidly collapse and form stars, since the pressure exerted by the particles of a
$10^4$~K, \lya-emitting gas would be insufficient to balance the effect
of gravity. This scenario is not consistent with our
observations (Section \ref{sec:powering}). On the other hand, the
\lya\ nebula may be globally at rest at the bottom of
the potential well if structured in
smaller and denser clouds moving with a typical velocity comparable
with the velocity dispersion of the cluster. However, the \lya\ clouds would dissipate
energy through turbulence. If not energized by external factors,
they would inevitably start
cooling and collapsing. All things considered, if
the \lya\ nebula were globally at rest in the dark matter halo, it would
quickly disappear on a cooling timescale, making our
discovery an unconvincingly lucky coincidence. Planned spectroscopic
follow-ups will directly probe the dynamical state of the nebula and
test our predictions.\\
\subsubsection{Cooling and heating}
Absent a strong powering mechanism, the continuous
irradiation of \lya\ photons would lead to the quick collapse and
disappearance of the clouds. This would happen on timescales of $t_{\mathrm{cool}} \, \thickapprox
2.07\times10^{11}\,\mathrm{s}\,(T/10^4\,\mathrm{K})
    (n_{\mathrm{e}}/1\,\mathrm{cm}^{-3})^{-1}
    \times(\Lambda(T)/10^{-23}\,\mathrm{erg}\,\mathrm{cm}^3\,\mathrm{s}^{-1})^{-1}
    \sim 0.1$~Myr, where $T=10^{4}$~K is the gas temperature,
    $n_{\mathrm{e}}\sim1-10$ cm$^{-3}$ the electron
density corresponding to plausible values of the volume filling factor
($f=10^{-3}-10^{-5}$), and $\Lambda(T)$ the cooling function \citep{dey_2005,
  sutherland_1993}. Strong cooling of the \lya\ clouds is disfavored
by the large extension of the nebula and the absence of features of
recent star-formation occurring in the
ICM (Section \ref{sec:powering}). Moreover, the cold gas is immersed
in a bath of energetic photons produced by the AGN that can keep a
large fraction of it ionized. This would be compatible with the geometry of
the system and dust absorption (Section
\ref{sec:agn}). In addition, magnetic fields in the ICM can insulate and stabilize
the ionized clouds, further preventing cooling and prolonging their
lifetime up to $\sim10$~Myr, as proposed for nebular
filaments in local CCs \citep{conselice_2001,
  fabian_2003, fabian_2008}.\\
Conversely, \lya-emitting gas clouds in macroscopic motion
with respect to the hot medium, can be thermalized through
hydrodynamical instabilities and shocks. We estimate the timescale
for the interaction between the cold and hot ICM phases following
\cite{klein_1994}:
\begin{equation}
t_{\mathrm{therm}} =
\left(\frac{n_{\mathrm{e}}^{\mathrm{Ly\alpha}}}{n_{\mathrm{e}}^{\mathrm{hot}}}\right)^{1/2}\,\frac{R_{\mathrm{cloud}}}{c_{\mathrm{s}}^{\mathrm{hot}}}
\end{equation}
where $R_{\mathrm{cloud}}$ is the radius of individual \lya-emitting clouds, and
$c_{\mathrm{s}}^{\mathrm{hot}} \approx 500$ \kms\ the sound speed in
the hot medium. This speed is also comparable with the velocity
dispersion in the cluster (Section \ref{sec:halo}). For simplicity, we
adopted the classical hydrodynamical, non-radiative case where we
considered the effects of hot winds moving at a typical speed of the
order of $c_{\mathrm{s}}^{\mathrm{hot}}$, much greater than the sound
speed in the cold gas. However, a fully numerical treatment including
radiative losses gives similar results \citep{scannapieco_2015}. We
allowed for possible clumpiness in the nebula
assuming $R_{\mathrm{cloud}} < R_{\mathrm{neb}}$ when the volume
filling factor is $f<1$, where $R_{\mathrm{neb}}=46$~kpc is the
  radius of the whole nebula (Section \ref{sec:lya_mass}). To constrain
$R_{\mathrm{cloud}}$, we adopted a pure geometrical approach \citep{hennawi_2013}. Assuming spherical clumps spatially
uniformly distributed in the spherical nebula of radius $R_{\mathrm{neb}}$ and with a single
uniform clumps' gas density, we can link $f$ to
$R_{\mathrm{cloud}}$ through the covering factor $f_{\mathrm{C}}$:
\begin{equation}
f = f_{\mathrm{C}}\frac{R_{\mathrm{cloud}}}{R_{\mathrm{neb}}}
\label{eq:fc}
\end{equation}
\begin{figure}
  \includegraphics[width=\columnwidth]{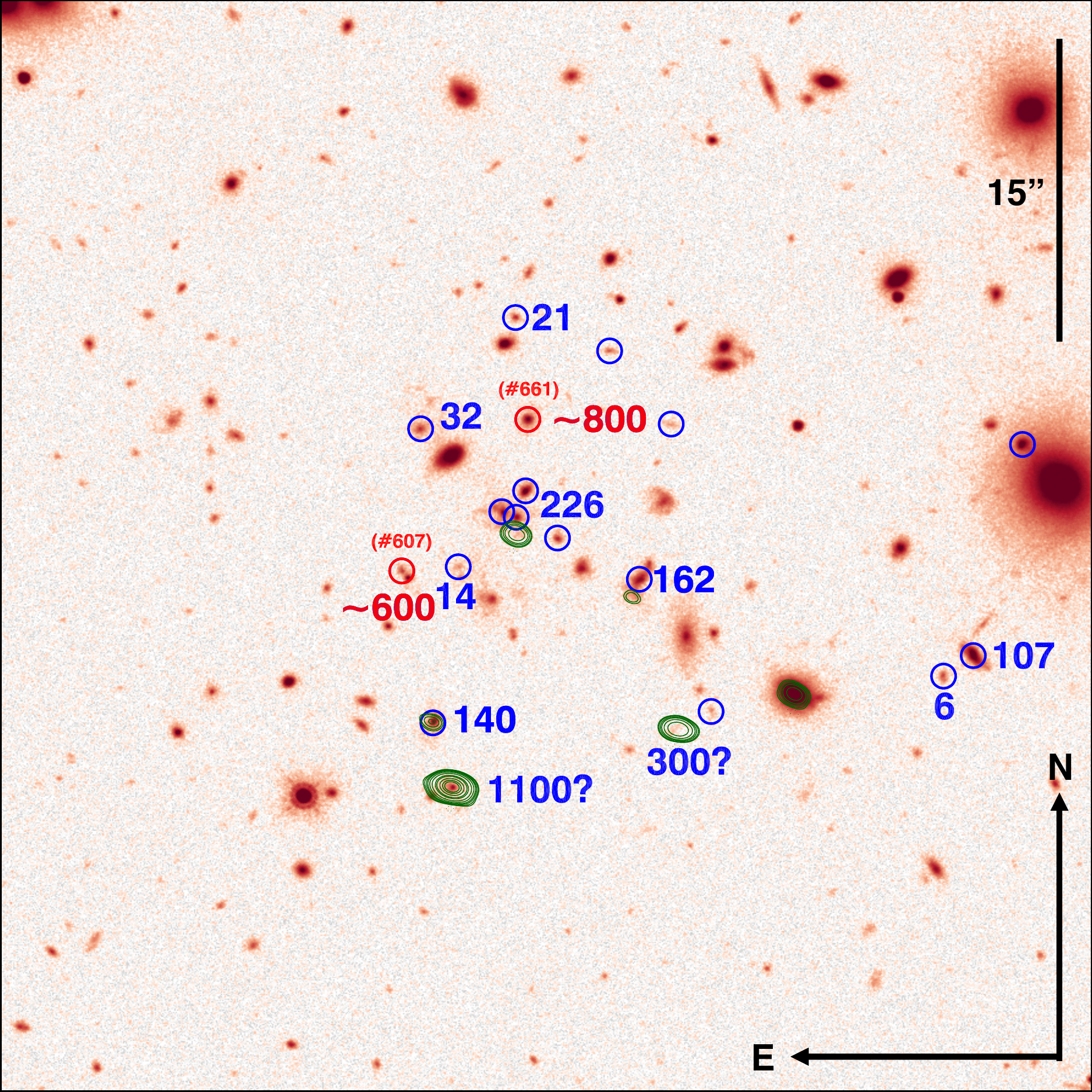}
  \caption{\textbf{Map of the activity in the central cluster region.}
  For each cluster star-forming (blue circles) or AGN (red circles)
  member in the \textit{HST}/WFC3 F140W
  image, we indicate the indirect estimates of the outflow rate in M$_{\odot}$
  yr$^{-1}$ as derived in Section \ref{sec:activity}. When available, we
  provide the SFR estimated from the continuum at 870 $\mu$m from ALMA
  observations (green contours). Question marks denote outflow rates
  associated to ALMA sources without redshift confirmation and hence not
  included in our most conservative approach.}
\label{fig:activity}
\end{figure}
The relative smooth
  morphology of the nebula and the flat surface brightness profile
  are consistent with $f_{\rm C}$ of the order of unity, although we
  cannot determine its accurate value. Assuming $f=10^{-3}-10^{-5}$, we obtain a typical timescale for
thermalization of $t_{\rm therm}\sim0.1-3$~Myr. Relaxing the constraint on the covering
factor up to a factor of $5$, we find $t_{\mathrm{therm}}\sim0.5-10$~Myr, consistent with
the lower limit on the lifetime of filaments in local CCs.\\
As a direct consequence, barring an
improbable observational coincidence, maintaining
the nebula stable against evaporation requires a replenishment of cold gas at a rate 
of $\dot{M}_{\mathrm{repl}} =
M_{\mathrm{Ly\alpha}}/t_{\mathrm{therm}} \gtrsim 1000$ \myr.
Note that this estimate is sensitive to the presence of colder gas reservoirs not
  shining in \lya\ and possible localized cooling partially compensating the
  heating, which could lower the final value. On the other hand, the
  quoted number
  could be regarded as a lower limit, since the parameters in the equations could
  substantially increase the replenishment rate in the plausible ranges we
  considered. The replenishment
rate is directly proportional to $f_{\rm C}$, but mildly depends on $f$
through both terms of the fraction ($\dot{M}_{\mathrm{repl}}\propto
f^{-0.25}$), making the minimum replenishment stable against the
range of values we allowed for the filling factor. Physically, the
smaller the volume filling factor, the smaller the total mass of the
nebula, but the shorter the evaporation time of the denser and clumpier
gas. The density contrast term and the size of the clumps act in
opposite way on $t_{\mathrm{therm}}$, with $R_{\mathrm{cloud}}$
dominating the final value: smaller clumps are
crossed by shocks or hydrodynamical perturbations more rapidly than
larger clouds and, consequently, they are disrupted
faster.\\

If not continuously sustained against evaporation, the nebula would
disappear on timescales of $t_{\mathrm{therm}}$ or, analogously, very
short timescales imply unphysical replenishment rates
$\dot{M}_{\mathrm{repl}}$ to explain the presence of the nebula.   
We note that the evaporation timescale is shorter than the
nebula crossing time ($\sim90$~Myr), given a radius of $R=46$~kpc and
a typical wind speed of $500$ \kms\ \citep[i.e.,][]{forster-schreiber_2014}. This raises the
problem of explaining the extension of the nebula, since the
\lya-emitting clouds should evaporate well before filling the observed
volume. The issue would be naturally fixed if the clouds primarily form \textit{in
  situ} by cooling from the X-ray emitting ICM. Globally, this is unlikely to be
the case especially far away from the cluster center, where cooling
times from bremsstrahlung are long. However, \textit{local}
thermal instabilities might be established in the densest portions of the ICM, providing part of the cold gas
needed. On the other hand, if the gas
replenishment is due to galaxies (as we envisage in the next section),
the size of the nebula is explained both by the
distribution of cluster members over a large area, since in this case
clouds being injected at different positions would not need to cross
the whole nebula, and by recent models of radiatively
cooling winds \citep{thompson_2015}. Moreover, galaxies are rapidly
moving in the cluster core and, consequently, winds are naturally
spread over large portions of the nebula.

\section{Discussion}
\label{sec:discussion}
In the previous section we have shown that the nebula must be
  constantly replenished of cold gas at a rate of $\gtrsim1000$ \myr\
  in order to shine for timescales longer than $\approx10$ Myr. Here
  we focus on
galaxy outflows as a plausible mechanism to supply this gas. We
introduce independent constraints on the amount of gas
released by galaxies based on the observed star formation and AGN
activities and we show that outflows are sufficient to explain the presence of the nebula.
We further discuss the implications of mass and energy
extraction from galaxies and the ensuing injection into the ICM, a
process necessary to explain the thermodynamics of local clusters. We
also draw a comparison with state-of-the-art cosmological simulations
to test the consistency of our estimate. 

\subsection{Gas replenishment through galaxy outflows}
\label{sec:activity} 
\begin{figure*}
  \centering
  \includegraphics[width=\textwidth]{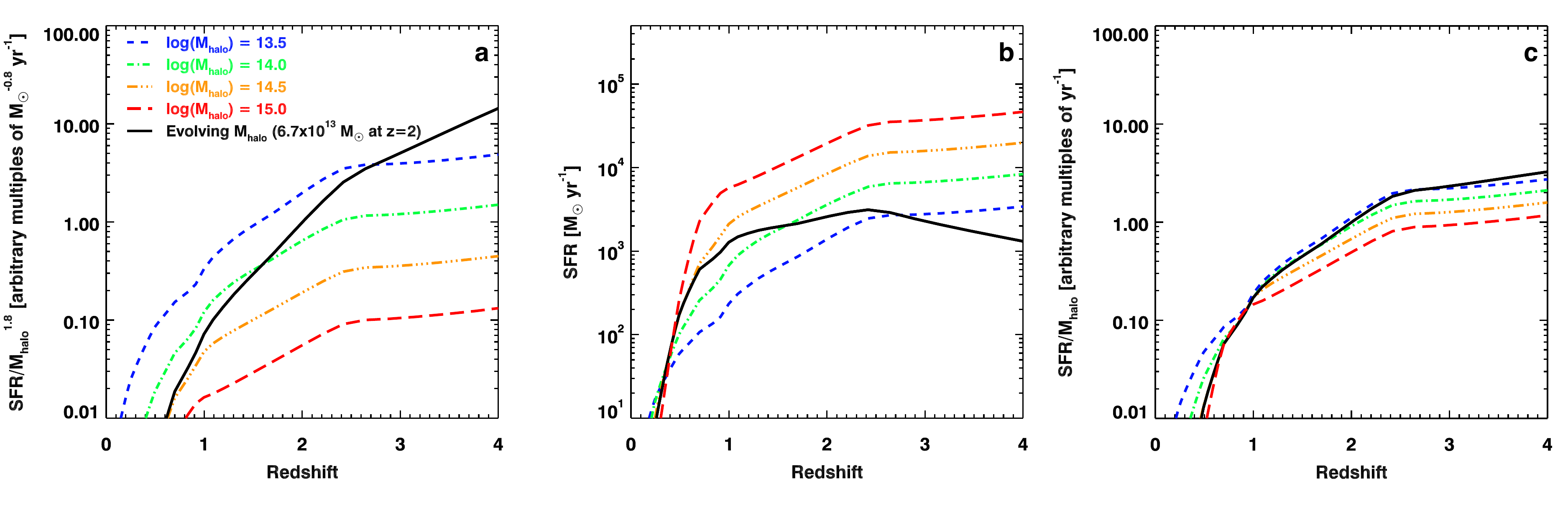}
  \caption{\textbf{Expected efficacy of outflow energy injection as a function of halo mass and redshift.}
    Based on the empirical mapping of the star formation and galaxy
    clustering evolution through cosmic time \citep{bethermin_2013},
    we model the redshift evolution of the outflow energy injection over
    the thermal energy of the ICM (panel \textbf{a}). The mechanical energy injection scales as
    the integrated SFR in the halo, while the total thermal energy of the hot ICM increases as
    $E_{\rm therm}\propto T_{\rm vir}\, M_{\rm gas} \propto
    M_{\mathrm{halo}}^{1.8}$, assuming a gas fraction varying with the
    halo mass \citep{renzini_2014}. Hence, the y axis in panel \textbf{a}
    represents the `efficacy' of the energy injection. In panel
    \textbf{b} we show the evolution of the SFR as a function of
    redshift and halo mass in \cite{bethermin_2013}. The ratio
    SFR$/M_{\mathrm{halo}}$ changes slowly between $2<z<4$ (panel \textbf{c}).}
\label{fig:matthieu}
\end{figure*}
The gas necessary to sustain the \lya\ nebula can be supplied by galaxy members
through supernovae- (SNe) and AGN-driven outflows, a feature
ubiquitously observed in high-redshift galaxies \citep[i.e.,][]{newman_2012,
forster-schreiber_2014, genzel_2014, harrison_2015} and strongly
supported by theoretical models and cosmological and zoom-in
simulations \citep[i.e.,][]{dave_2008, hopkins_2012, lilly_2013,
  gabor_2014}. Is the galaxy activity in
CL~J1449+0856's core sufficient to provide a
minimal mass rate of $\gtrsim1000$ \myr\ as required by the \lya\
nebula? To answer this question, we computed the total mass outflow
rate considering both contributions from the observed SFR and AGN
activity (Figure \ref{fig:activity}). We converted members' SFRs into
mass outflow rates $\dot{M}_{\mathrm{out}}$
by multiplying by a conservative mass loading factor $\eta =
\dot{M}_{\mathrm{out}}/\mathrm{SFR}= 1$. This is likely to be a
lower limit for the ionized and molecular gas expelled by galaxies,
both observationally and theoretically \citep[i.e.,][Hayword et al. 2015]{newman_2012, hopkins_2012,
gabor_2014}. This order of magnitude is also necessary to explain the
metal enrichment of the ICM. Indeed, the same amount of metals
is locked into stars and distributed in
the ICM, favoring the equality
$\dot{M}_{\mathrm{out}}\approx\mathrm{SFR}$ \citep[i.e.,][]{renzini_2014}. The SFRs were derived either
from SED modeling from our 13-band
photometry (S13), H$\alpha$ from our recent Subaru/MOIRCS follow-up
\citep{valentino_2015}, or
870 $\mu$m continuum detection in ALMA maps applying a Main-Sequence
galaxy template \citep{magdis_2012}. ALMA observations, reduction and analysis will be
presented in a forthcoming paper (Strazzullo et al in prep.). The total
SFR in the central region is $\mathrm{SFR}\approx1000$
\myr.
An individual bright ALMA source stands out in the cluster field.
Its 870 $\mu$m flux is $F_{\mathrm{870\mu m}} = 5.5$~mJy, corresponding to a total
infra-red luminosity between $8-1000$~$\mu$m of
$L_{\mathrm{IR}} = 6.6\times10^{12}$ L$_\odot$ and $\mathrm{SFR} = 1100$ \myr\ at
$z=1.99$ (Figure
\ref{fig:activity}). The measurement errors are negligible with
respect to the $0.15$ dex uncertainty due to modeling (Strazzullo et
al. in prep.). As there is no spectroscopic confirmation that the ALMA
source is a member of the cluster, we have conservatively excluded it
from the SFR accounting. We note that, if confirmed to be
part of the cluster, this source would increase by a factor of $2\times$ the total
SFR in the core.\\
The growth of black holes further contributes to the mass
  outflow rates. We estimated its order of magnitude by directly converting
$L_{\mathrm{bol}}$ into mass outflow rates using the empirical
calibration by \cite{cicone_2014}. In this case, we obtain $\approx600$ and $800$ \myr\
for \#607 and \#661, respectively. Moreover, it appears that we have not captured the system 
during a phase of exceptional AGN activity. In fact, the integrated SFR/$L_{\rm X}$
ratio observed in the cluster core is close to the cosmic average
value \citep{mullaney_2012}. The predicted X-ray luminosity is 
$\langle{L}_{\rm X} \rangle = \mathrm{SFR}\times4.46\times10^{41}$ \es\ $\approx
4.5\times10^{44}$ \es, while the
observed value from the two AGN is $3.4 \times10^{44}$ \es.
We remark that the calibration by \cite{cicone_2014} is based on a
sample of local bright IR galaxies with previously known outflows, which, in
principle, may overestimate the outflow rates if the relation captures
a phase shorter than the AGN duty cycle. On the other side,
contribution from phases other than molecular and the
uncertain CO luminosity-to-gas mass conversion can increase the
outflow rates derived with this calibration. Indeed, strong
nuclear ionized winds are now observed in fractions up to 50-70\% of high-redshift AGN
\citep{harrison_2015}, showing how common these features are. Moreover,
the calibration by \cite{cicone_2014} is in line with the expectations from simulations
reproducing the relations among black
hole and galaxy bulge masses or velocity dispersions. In terms of the ratio between the kinematic
energy released by AGN per unit time and their bolometric luminosity, simulations
usually assume a coupling efficiency $\epsilon_{\rm f}\sim0.05-0.15$
\citep[i.e.,][and Section \ref{sec:simulations} below]{dimatteo_2005,
  lebrun_2014}. As we show in Section \ref{sec:instantaneous_injection},
the instantaneous kinetic energy associated with AGN and mass outflow
rates estimated from the \cite{cicone_2014} relation is indeed $\sim5$\% of the
observed bolometric luminosities. Therefore, all things considered, we 
do include an AGN contribution following \cite{cicone_2014} in our
fiducial estimate of the total mass outflow rate.\\
 Finally, we note that the
  reasonable agreement between the replenishment rate estimates from the galaxy
  activity in the core and from the \lya\ nebula would be just incidental if the \lya\ emission were
predominantly powered by shocks induced by galaxy outflows on the surrounding
pressurized ICM (see Section \ref{sec:shocks}), suggesting a lesser contribution from this
mechanism. In this case, the estimate of the
replenishment rate reported in Section \ref{sec:timescales} would not
be valid. However, the independent constraint on the energy injection by galactic winds
presented in the following section would be unaffected.

\subsection{Energy injection into the ICM}
\label{sec:injection}
Together with mass, outflows extract energy from galaxies and then
deposit it into the surrounding ICM through dissipation, shocks or
turbulence. In the following sections we estimate the \textit{kinetic}
energy injection, neglecting alternative contributions, i.e. from
radiation.

\subsubsection{Instantaneous injection}
\label{sec:instantaneous_injection}
First, we can estimate the
\textit{instantaneous} injection of energy at the time of observation:
\begin{equation}
\dot{E}_{\mathrm{kin}} = \frac{1}{2} \dot{M}_{\mathrm{out}} v^2
\label{eq:einst}
\end{equation}
where $\dot{M}_{\mathrm{out}}$ is the total amount of gas ejected per unit
time at $z=1.99$ by galaxies and $v$
is the outflow velocity. We do not measure $v$ in individual
members in our sample, but its statistical average is quite well
constrained by increasing samples of high-redshift
observations. Therefore, our estimate of $\dot{E}_{\mathrm{kin}}$ should be taken in a
statistical sense. We assign a wind speed of 500 \kms\ to SN-driven
outflows for each star-forming galaxy, while for AGN-driven outflows we assume a typical speed
of $1000$ \kms\ \citep{genzel_2014, forster-schreiber_2014,
cicone_2014}. Given the baseline mass outflow rate in Section
\ref{sec:activity}, we obtain $\dot{E}_{\mathrm{kin}}(z=1.99) \sim 5 \times 10^{44}$ \es. This energy
is a factor of $20\times\,(5\times)$ larger than the observed \lya\ (X-ray)
extended luminosity. The $5\times$ factor with respect to the X-ray
luminosity is sufficient to offset the global radiative cooling of the
hot plasma. Assuming the balance between heating and the observed
cooling rate as in local clusters would thus imply an energy injection
$5\times$ lower than estimated above. However, net heating is
necessary to justify the presence of the \lya\ nebula, since the
cooling from the X-ray globally occurs on long timescales and is not
sufficient to explain the \lya\ emission (Section \ref{sec:coolcores}).  
The injected energy is coming predominantly from AGN activity ($\sim 85$\%)
with a non-negligible contribution from star formation
($\sim 15$\%), while, in terms of mass, AGN are responsible for up to
$2/3$ of the total gas released into the ICM. SFGs would dominate the
mass and energy injection only if we largely overestimated the contribution
from AGN.
We note that CL~J1449+0856 is not anomalous in terms of
star formation activity with respect to potentially similar
structures at comparable redshift \citep[i.e.,][]{tran_2010, yuan_2014,
santos_2015, gobat_2015} and it is
globally consistent with the tracks reported in Figure
\ref{fig:matthieu} based on the model by \cite{bethermin_2013}. The instantaneous
energy input from AGN corresponds 
only to $0.05\,L_{\rm bol}^{\rm{AGN}}$, a factor of $3\times$ lower than
typically assumed in simulations (Section \ref{sec:simulations}),
supporting the estimate of the mass outflow rates reported in Section \ref{sec:activity},
while from star formation it is just $0.003\, L_{\rm bol}^{\rm SFGs}$. In general, given the
SFR/$L_{\mathrm{X}}$ cosmic average \citep{mullaney_2012} and the
adopted calibrations, we expect
AGN outflows to provide $5-10\times$ more energy than winds induced by
star formation.

\subsubsection{Integrated energy injection}
We can now estimate the \textit{total} energy
injection up to $z=1.99$, integrating $\dot{E}_{\mathrm{kin}}$ over time prior to
observation: 
\begin{equation}
E_{\mathrm{kin}} = \int_{t(z\geq1.99)} \dot{E}_{\mathrm{kin}}\,dt
\label{eq:etot}
\end{equation}
For simplicity, we assume that the \textit{instantaneous} energy
injection is proportional to the SFR:
\begin{equation}
\dot{E}_{\mathrm{kin}} = \beta\,\mathrm{SFR}
\label{eq:beta}
\end{equation}
where $\beta(z=1.99) \sim 1.6\times10^{49}$ erg \msun$^{-1}$.
Then from Eq. \ref{eq:etot}:
\begin{equation}
  \begin{aligned}
    E_{\mathrm{kin}} = & \int_{t(z\geq1.99)} \beta\,\mathrm{SFR}(t)\,dt \\ 
                          = & \frac{\beta}{1-R} \int_{t(z\geq1.99)}
                          \mathrm{SFR}(t) (1-R)\,dt \\
                          = & \frac{\beta}{1-R}\,M_{\star}
    \end{aligned}
\end{equation}
where $M_{\star}$ is the total stellar mass of $2\times10^{12}$ \msun\
observed at $z=1.99$ (Section \ref{sec:halo} and S13) and $R=0.4$
is the mass return into the interstellar medium
\citep{bruzual_2003}. Eventually, we obtain $E_{\mathrm{kin}} =
5\times10^{61}$ erg. Considering a universal baryon fraction of
$f_{\rm b}=0.15$ in the ICM \citep{planck_2014}, the total energy
per particle in the hot ICM then is $\sim2$ keV. This value is $\sim10$\% of the
  binding energy of the halo at $z=1.99$ and of the same order of
  magnitude in cluster progenitors. Hence, part of the ICM particles might
  have been expelled by the structure at some early stage. The integrated
  energy is also comparable with the thermal energy per
particle $E_{\mathrm{therm}} = 3/2\,k\,T$. Indeed, assuming
virialization, $kT=kT_{\mathrm{vir}} = GM_{\mathrm{halo}}\mu
m_{\mathrm{H}}/2R_{\mathrm{vir}} \sim 1.9$ keV and, thus, $E_{\mathrm{therm}} \sim
2.8$ keV. This is an order of magnitude estimate, as the structure is
unlikely to be fully virialized at this stage -- simulations suggest a
thermodynamic temperature $15-20$\% smaller than $T_{\rm vir}$
(Section \ref{sec:simulations}). We stress here that
  our estimate of the integrated energy injection is affected by
  uncertainties on the total mass outflow rate, outflow velocities,
  the halo mass, and its baryon content and it depends on the assumptions we
  described. All things considered, the estimate may well increase or
  decrease by a factor of $\sim0.5$~dex.\\ 
This approach relies on the use of $M_\star$ in
CL~J1449+0856 as a proxy for the total
mass ejected through outflows in the past. This presumes the adoption
of a mass loading factor of $\eta=1$ and that
$v$ depends on local galaxy properties not evolving with
time. The advantage of using $M_{\star}$ is the straightforward
inclusion of the contribution to the energy injection by galaxies active in the past,
but observed to be passive at $z=1.99$. However, there are two important
assumptions behind this results: first, we suppose that the
total AGN mass outflow rate is proportional to the total SFR at any
time and second, that $\beta$ is constant with time. 

\subsubsection{Caveats}
We justify the first assumption considering that SFR and AGN activity are correlated
\citep{mullaney_2012}: statistically, on large samples the average AGN X-ray luminosity
is equal to $\langle L_{\rm X} \rangle = \mathrm{SFR}\times4.46\times10^{41}$ \es. Nevertheless, the AGN mass
outflow rate might depend non-linearly on the AGN luminosity. For
example, in the empirical relation by \cite{cicone_2014},
$\dot{M}_{\mathrm{out}} \propto L_{\mathrm{bol}}^{\mathrm{b}}$ with $\mathrm{b}=0.72$. From
Eq. \ref{eq:einst}, this non-linear term becomes:
\begin{equation}
    \dot{E}_{\mathrm{kin}}^{\mathrm{AGN}} = \frac{1}{2} \dot{M}
    _{\mathrm{out}} ^{\mathrm{AGN}} v_{\mathrm{AGN}}^2
     =  k_{1}\, L_{\mathrm{X}}^{0.72} 
      =  k_{2} \,\mathrm{SFR}^{0.72}
\label{eq:nonlin}
\end{equation}  
where $k_{1}$ and $k_{2}$ are constants including the bolometric
correction linking $L_{\mathrm{X}}$ and $L_{\mathrm{bol}}$, the
velocity term $v^2/2$,
and the coefficients in the \cite{cicone_2014} and
\cite{mullaney_2012} relations. Simply combining Eq. \ref{eq:beta} and Eq. \ref{eq:nonlin},
we obtain:
\begin{equation}
    \beta = c_1 + c_2\,\mathrm{SFR}^{0.72-1}
    \label{eq:final}
\end{equation}  
where $c_1$ and $c_2$ are constants. Thus, the non-linear term
introduced by the AGN mass outflow rate impacts our result only
when the total SFR in the progenitors of CL~J1449+0856 drops significantly.   
Eq. \ref{eq:final} justifies also our second main assumption that $\beta$
is roughly constant with time, depending only on numeric constants and the total SFR
in all the cluster progenitors.\\
Does the total SFR in the cluster progenitors evolve with redshift?
At $z>1.99$ the SFR is spread over several subhalos that will form
the observed cluster by merging. Here we trace the growth of individual
dark matter halos from simulations using the \cite{fakhouri_2010}
model. According to \cite{bethermin_2013}, in each
subhalo the total SFR peaks at $z\sim2$ and then slowly
decreases (black curve in Figure \ref{fig:matthieu}, panel
b). However, to compute the total SFR
contributing to the energy injection over time we have to consider \textit{all} the
subhalos. This corresponds to normalizing the individual SFR to the
halo mass at each redshift (Figure \ref{fig:matthieu}, panel c). In
this case, the function $X(z) = \langle \mathrm{SFR}(z) \rangle /
M_{\rm halo}(z)$ mildly increases with redshift. Thus, the non-linear
term in Eq. \ref{eq:final} becomes less important with redshift.

\subsubsection{Final remarks}
We attempted an alternative estimate of the total kinetic energy
purely base on the tracks in Figure \ref{fig:matthieu}. We obtain
$E_{\mathrm{kin}} \sim 5 \times10^{61}$ erg released
by galaxies over $2<z<4$ computed as:
\begin{equation}
    E_{\mathrm{kin}} = \frac{1}{2}\dot{M}_{\mathrm{repl}} v^2
    \int_{t(z=4)}^{t(z=2)} \! \frac{X(t(z))}{X(t(z=2))} \, \mathrm{d}t
\end{equation} 
where the function $X(z) = \langle \mathrm{SFR}(z) \rangle / M_{\rm halo}(z)$
accounts for the expected flat trend of $\langle \mathrm{SFR} \rangle$ at
$2<z<4$ and incorporates the integrated activity spread in halo progenitors of
lower masses (Figure \ref{fig:matthieu}, panel c). The net effect of the integral is an increase
of the time interval, from 1.7~Gyr between $2<z<4$ to 4.4~Gyr. This
result is consistent with the one presented above, providing $\sim2$
keV per particle in the hot ICM, assuming a universal baryon fraction
$f_{\rm b}=\Omega_{\rm
  b}/\Omega_{\rm m} =0.15$.\\ 
Here we limit the integral to $z=4$, before which the masses of individual
progenitor halos rapidly become similar to individual
galaxy halos ($\approx1\times10^{13}$ M$_\odot$ following
\cite{fakhouri_2010}). At these masses, fast winds would have easily
expelled the material from the halo, that later would have been
reaccreted with the halo growth. However, observed properties of local
structures may disfavor this scenario for energy injection \citep{ponman_2003}.\\
 We note that the tracks in Figure \ref{fig:matthieu} are calibrated on
 the observed stellar mass function of passive and star-forming
galaxies residing in halos of masses of $11.5 < M_{\mathrm{halo}} <
13.5$ at high redshift. However, the model does not assume any environmental
dependence of galaxy properties, prominent at lower
redshift. The transformation of \textit{cluster} galaxies
into red, passive, early-type systems at low redshift makes the
predicted SFR a likely overestimation at $z\lesssim1.5$ \citep{popesso_2015}. 
Below $z\sim1.5$ the outflow energy contribution to the ICM is
expected to be negligible with respect to the internal energy, as
shown in Figure \ref{fig:matthieu}. We remark here that we do not make
any prediction on the later growth of a massive central galaxy and its
associated black hole, whose ``radio'' maintenance feedback looks necessary to
avoid overcooling in the cluster core \citep{mcnamara_2007,
  fabian_2012, gaspari_2012}. 

\subsection{Comparison with cosmological simulations}
\label{sec:simulations}
\begin{figure*}
  \centering
  \includegraphics[width=\textwidth]{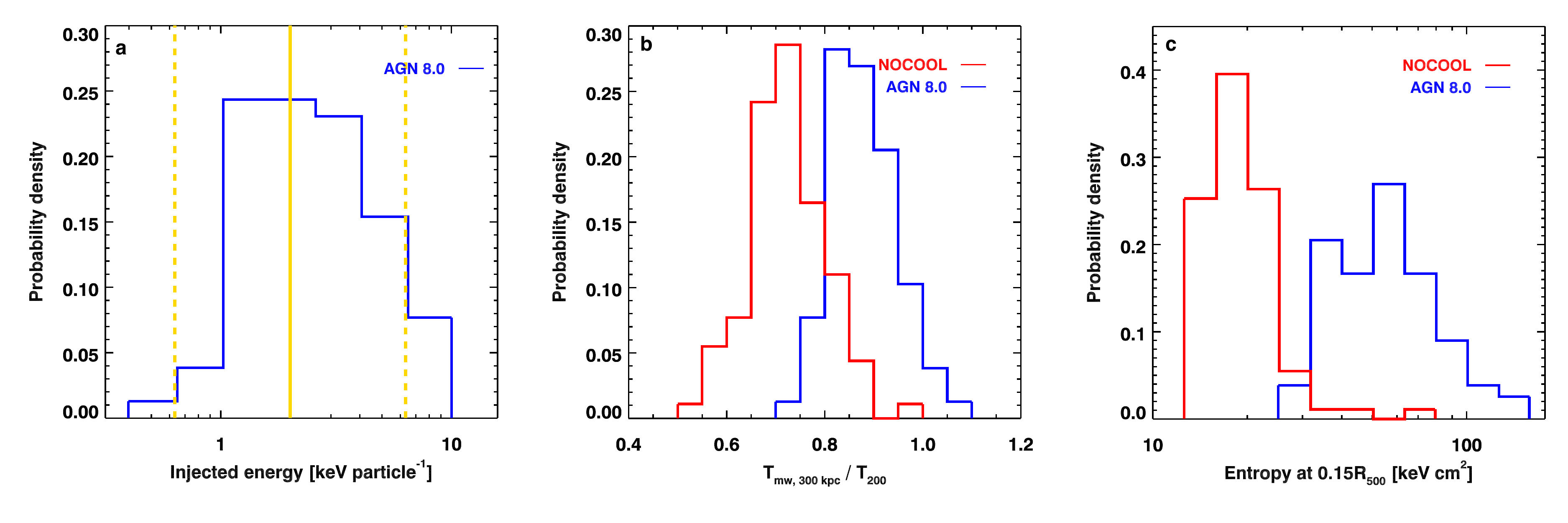}
  \caption{\textbf{Temperature and entropy increase in $z=2$ clusters when AGN
      feedback is active in simulations.} We show the distribution of the
    energy per particle injected in $M_{200}= (5 - 7)\times10^{13}$
    M$_{\odot}$ systems at $z=2$ when AGN feedback is turned on in our
    suite of cosmo-OWLS simulations (panel \textbf{a}) and its effect
    on the mass-weighted temperature within 300 kpc in units of the virial
    temperature $kT_{\rm vir}$ (panel \textbf{b}) and entropy at
    $0.15R_{\mathrm{vir}}$ (panel \textbf{c}). Blue lines indicate the
    reference AGN 8.0 feedback model, while red lines mark the
    non-radiative model \citep{lebrun_2014}. The yellow lines in
    panel \textbf{a} show our fiducial estimate of $\sim2$ keV per particle
    from observations and a 0.5~dex uncertainty.}
  \label{fig:sim}
\end{figure*}
We compared our observational results with the total energy injected
by black holes into the ICM of
systems similar to CL~J1449+0856 at $z=2$ in simulations. We used the
two models from the suite of hydrodynamical cosmological
simulations presented in Le Brun et al. (2014), which form
an extension to the Overwhelmingly Large Simulations project \citep[OWLS,][]{schaye_2010}. 
The first is a standard non-radiative model (NOCOOL), while the second
further includes prescriptions for metal-dependent radiative cooling, star formation,
stellar evolution, mass loss, chemical enrichment, stellar feedback, and AGN
feedback. Among the models described in \cite{lebrun_2014}, we selected the AGN 8.0 model as it provides the best
match to the X-ray, Sunyaev-Zel'dovich, and optical observations of
local groups and clusters \citep{lebrun_2014, mccarthy_2014}. In these two models, we selected the halos with
$M_{200}= (5 - 7)\times10^{13}$ M$_{\odot}$ at $z=2$ (yielding respectively 79 and 91 such systems
in the AGN 8.0 and NOCOOL physical models). For each of these structures,
we computed the mass-weighted temperature within a 300 kpc aperture,
the mean entropy $S=kT/n_{\mathrm{e}}^{2/3}$ within $0.15\,R_{500}$, the virial temperature
$kT_{\rm vir}$, and the binding energy. The energy injected by
all the black holes lying within $R_{500}$ is
$E_{\rm inj}=M_{\rm BH}(<R_{500})c^{2}\epsilon_{\rm f}\epsilon_{\rm r}/(1-\epsilon_{\rm
  r})$, where $\epsilon_{\rm r} = 0.1$
is the radiative efficiency of the black hole accretion disk,
$\epsilon_{\rm f} = 0.15$ the efficiency of the coupling of the AGN feedback to the
gas, and $c$ the speed of light. We estimate the average injected
energy per particle assuming $f_{\rm b}M_{500}/\mu m_{\rm p}$ baryonic
particles in the ICM, where $f_{\rm b}=0.15$ is the universal
baryon fraction \citep{planck_2014}, $\mu=0.6$
is the mean molecular weight, and $m_{\rm p}$ the proton mass.  
We obtain that the mean injected energy is of the order of $8\times10^{61}$
erg ($\approx2.8$ keV per particle, Figure
\ref{fig:sim}, panel a), which is of the same order of
magnitude as the typical binding energy of the selected
systems. Using $M_{200}$ instead of $M_{500}$ in the definition of the number of
particles reduces the estimate by a factor $1.4\times$. However, we
stress that this is a rough estimate of the overall effect on the
whole ICM, while in the simulations the energy injection is effective
mostly in a small region surrounding the AGN. All things considered,
this estimate is fully consistent with our
  observational estimate of $\sim2$ keV per
particle.
The mean temperature increases from $1.44$ keV to $1.73$ keV when
efficient AGN feedback is included (Figure
\ref{fig:sim}, panel b). Moreover, the
entropy within $0.15\,R_{500}$, tracing non-gravitational heating and
cooling, increases from $19.9$ keV cm$^2$  to $58.0$ keV cm$^2$ (Figure
\ref{fig:sim}, panel c). 
As the mean baryonic fraction within $R_{500}$
decreases from $14$\% 
in the non-radiative model to $10.7$\% in the AGN 8.0
model, some of the gas which should have been contained within $R_{500}$ in
the absence of AGN feedback has been ejected, similarly to what
was previously found for progenitors of $z=0$ groups (McCarthy et
al. 2011, but see \citealt{pike_2014} who find that most of the AGN
feedback energy is released at $z<1$ in their simulated clusters).
Overall this set of cosmological simulations predicts an
energy injection due to AGN of the same order of magnitude of
our estimate based on the average properties of galaxy outflows.

\subsection{Future \lya\ surveys of high-redshift clusters}
\label{sec:expectations}
The energy injection scenario based on galaxy outflows replenishing
huge gas reservoirs of cold and warm gas should apply for structures
similar to CL~J1449+0856 and comply with
the general increase of star formation and AGN activity observed in
high-redshift galaxies. Do we thus expect to see giant \lya\ nebulae in all
massive cluster progenitors? The answer could be negative. In fact,
AGN activity -- which illuminates the gas expelled through outflows and
keeps it ionized -- might be a prerequisite for the presence of \lya\
systems. Absent a
powerful ionizing source, dense environments hosting strong star
formation activity might
not show any extended \lya\ blob. This  might be the case for the
massive halo inside the proto-cluster region at $z=3.09$ in
the SSA22 field \citep{steidel_2000, kubo_2015}. A statistical
assessment of the number of active galaxies in clusters at each stage
of their evolution is important to address this issue. Nevertheless, galaxy
outflows remain an ubiquitous feature of high-redshift galaxies.
Are the massive gas reservoirs replenished by outflows destined
to collapse and form stars according to their cooling and free-fall
time? The gas in outflows is not at rest by
definition. Moreover both simulations
\citep{bournaud_2014} and observations \citep{martin_2009} show that
outflows accelerate at larger radii because of pressure gradients in
steady-state flows. This results in long collapse timescales, possibly
preventing the formation of stars spread over several tens of kpc. 
The assembly of larger samples of clusters progenitors will allow to
test these predictions.

\section{Conclusions}
\label{sec:conclusions}

In this work we presented the discovery of a giant $100$-kpc extended
\lya\ nebula in the core of a $5-7\times10^{13}$ \msun, X-ray detected cluster at
$z=1.99$. This discovery reveals the coexistence of
  warm ionized blobs and the hot intergalactic medium and extends the known relation between \lya\
nebulae and overdense regions of the Universe to the dense core of a
relatively mature cluster. We pinpointed two X-ray AGN as the most likely candidates to
power the nebula, disfavoring ionization from very young stars and
cooling from the X-ray phase in the form of a stationary classical
cooling flow. In principle, regulated cooling as in local cool-core
  clusters could partially contribute to the \lya\ luminosity, but
  several inconsistencies between CL~J1449+0856 and local systems are
  evident. Above all, the ratio between the \lya\ luminosity and the
  total X-ray luminosity is a factor $10-1000\times$ higher in
  CL~J1449+0856 than in local CCs even in those cases where strong
  radio-sources are present (i.e., Perseus). Dissipation of
    mechanical energy injected by galaxy outflows may also contribute
    to the total \lya\ luminosity. The interaction between the \lya\ nebula and the
surrounding hot ICM requires a $\gtrsim1000$ \myr\ gas replenishment
rate to sustain the nebula against evaporation. We explore
  galaxy outflows in the cluster core as a possible source of gas
  supply and find that the generous total SFR ($\approx1000$ \myr) and
  the outflow rate owing to the growth of supermassive black
holes ($\approx1400$ \myr) are sufficient to replenish
the nebula. This directly implies a
significant injection of kinetic energy into the ICM up to $\approx2$
keV per particle, in agreement with the predictions from the
cosmo-OWLS simulations and with constraints set by the
thermodynamic properties of local massive structures. In our
  baseline scenario the AGN channel
provides up to $85$\% of the total injected energy, with the rest
supplied by star formation through SNe-driven winds. The instantaneous
energy injection exceeds by a factor of $5$ the current X-ray
luminosity, offsetting the global cooling from the X-ray phase. Nevertheless, the
high star formation and black hole accretion rates deep in the
potential well of this cluster support the general increase in galaxy
activity observed in similar structures at comparable redshift and
challenge the current prescriptions on the fueling by cosmological cold
flows penetrating in massive halos. If this structure is not just a
curious anomaly, the potential presence of cold streams despite its high mass
would lead to important consequences on the ``halo quenching''
mechanism and, thus, on galaxy formation and evolution in general.\\

The advent of forthcoming facilities will allow us to drastically reduce
observational uncertainties and avoid a heavy resort to
assumptions. Measurements of temperature, pressure, and
entropy profiles of the hot ICM in young clusters will be possible
with the foreseen Athena X-ray satellite, while the systematic follow-up of \lya\
emission in clusters at $z>2-3$ could start soon with new wide field
integral field spectrographs on large telescopes, like MUSE and
KCWI. Spectroscopy in the ultra-violet range provides crucial
  information on the kinematics of the nebula, the metal enrichment, and
    its main powering mechanism. If the scenario we propose here is correct, we expect
  the \lya\ nebula to show signatures of complex motion due to
  outflows and to be fairly metal-rich. Eventually, the arising coherent scenario we sketch
could help to understand the global early evolution of
massive structures.

\acknowledgements
We acknowledge the constructive comments of the referee and
we thank Daniel Perley for the support during the
  reduction of Keck/LRIS data with his pipeline.
Some of the data presented therein were obtained at the W.M. Keck
Observatory, which is operated as a scientific partnership among the
California Institute of Technology, the University of California and
the National Aeronautics and Space Administration. Keck telescope time
was granted by NOAO (Prop. ID: 2014A-0131), through the
Telescope System Instrumentation Program (TSIP). TSIP is funded by
NSF. The Observatory was made possible by the generous financial
support of the W.M. Keck Foundation. The authors wish to recognize and
acknowledge the very significant cultural role and reverence that the
summit of Mauna Kea has always had within the indigenous Hawaiian
community.  We are most fortunate to have the opportunity to conduct
observations from this mountain. The scientific results reported in
this article also are based in part on observations made by the
Chandra X-ray Observatory. AF acknowledges the Chandra grant
GO4-15133A to UMBC. This paper also makes use of the following ALMA data:
ADS/JAO.ALMA\#2012.1.00885.S. ALMA is a partnership of ESO (representing
its member states), NSF (USA) and NINS (Japan), together with NRC
(Canada), NSC and ASIAA (Taiwan), and KASI (Republic of Korea), in
cooperation with the Republic of Chile. The Joint ALMA Observatory is
operated by ESO, AUI/NRAO and NAOJ.
We acknowledge financial support from
the Agence Nationale de la Recherche (contracts \#ANR-12-JS05-0008-01)
and the EC through the European Research Council Starting grants
StG-257720 and StG-240039, and the Advanced grant FP7-340519.

\bibliography{apj_lya_halo_bibliography}

\begin{figure}
  \centering
  \includegraphics[width=0.5\textwidth]{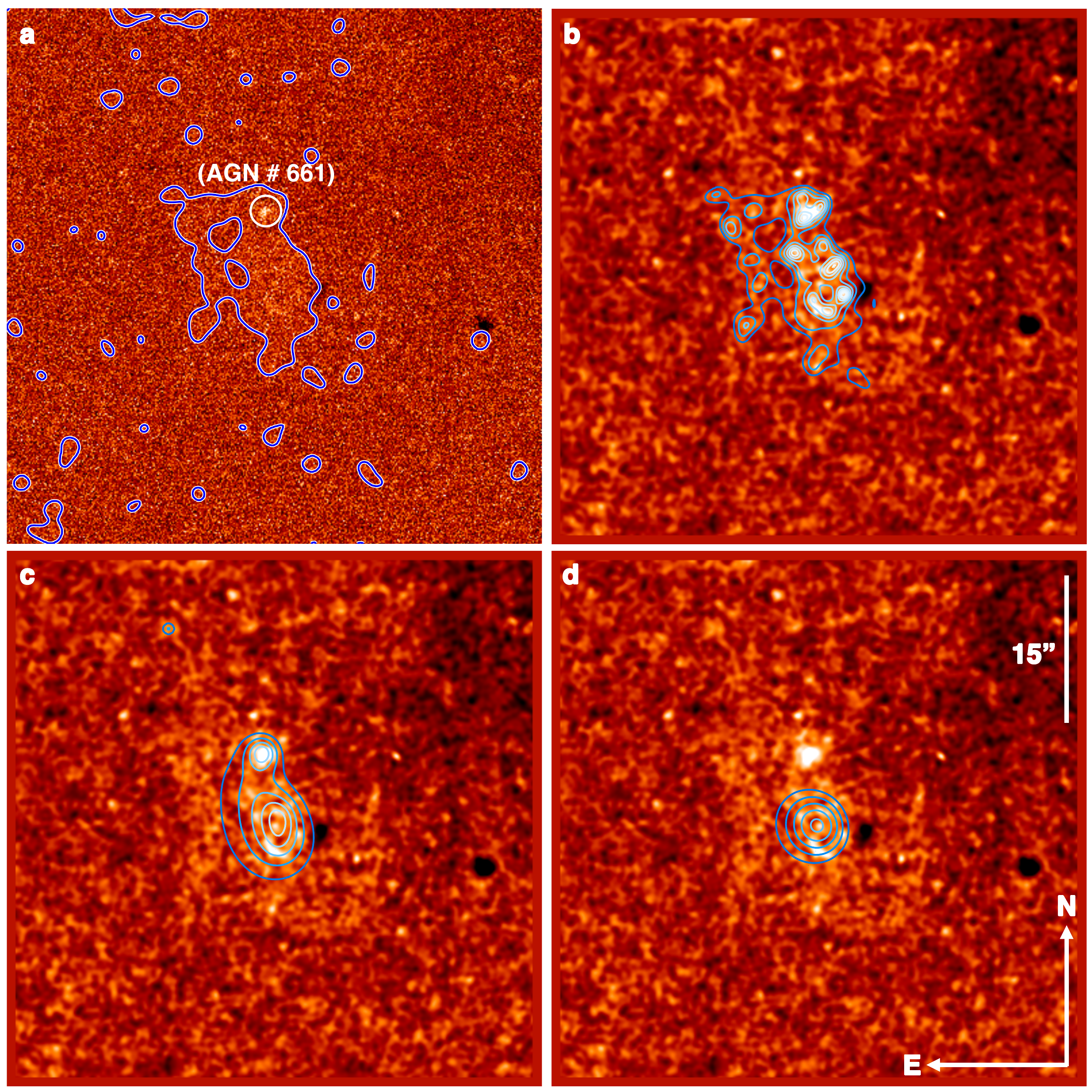}
  \caption{\textbf{Wavelet reconstruction of the \lya\ image.}
    We show the original unsmoothed \lya\ emission line map of the central region
    of CL~J1449+0856 in panel \textbf{a}. The $1\sigma$ contour of the large scale \lya\
    emission from the wavelet reconstruction (blue line) and the X-ray
    obscured AGN (white circle) are marked for reference. 
    Panels \textbf{b, c}, and \textbf{d} show the reconstructed
    wavelet contours at $\geq1\sigma$, $\geq3\sigma$, and
    $\geq5\sigma$ respectively (blue lines) of the \lya\ emission line
    map. Point-like sources have been subtracted before computing the
    surface brightness contours. The
    number of contours is arbitrary and chosen to pinpoint the peak of
    the extended emission. For reference, 15''
      correspond to $\sim125$~kpc at $z=1.99$.}
  \label{fig:appendix}
\end{figure}

\appendix
We show the unsmoothed \lya\ image from the Keck/LRIS narrow-band
follow-up of CL~J1449+0856 in panel (a) of Figure \ref{fig:appendix}. The
only purpose is to demonstrate that the \lya\ emission is not dominated by
individual galaxies, but it is distributed fairly homogeneously over
several square arcsec. The very low surface brightness regimes probed
in this image make the identification of the nebula difficult by
eye. It is easier to recognize it by comparing the original narrow- and
broad-band images shown in Figure \ref{fig:images} or, alternatively,
with a moderate smoothing (1'', Figure \ref{fig:images} and
\ref{fig:appendix} panels b-d). To guide the eye and pinpoint the peak of the
extended emission, in panels (b-d) of Figure \ref{fig:appendix} we show the contours of
the wavelet reconstructed \lya\ image. In each panel we show the contours after the subtraction of point-like sources, retaining only the signal on larger
scales, namely the \lya\ nebula. 
Panel (b) shows the maximum extension
of the \lya\ nebula, while the smoother denoised contours in panels (c) and (d)
allow for identifying the peak of the extended emission. The
appearance of two peaks in panel (c) depends on the number of scales adopted to
slice the image with the wavelet technique and does not affect the
main findings of this work. The region spanned by the $\geq5\sigma$
detection in panel (d) is the same used to measure the extended
continuum emission (Section \ref{sec:continuum}). In every panel the number of
contours is chosen arbitrarily to highlight the peak of the emission
and does not correspond to a fixed step in surface brightness.

\end{document}